\documentclass[sigconf]{acmart}

\AtBeginDocument{%
  \providecommand\BibTeX{{%
    \normalfont B\kern-0.5em{\scshape i\kern-0.25em b}\kern-0.8em\TeX}}}

\usepackage{enumitem}

\usepackage{dirtytalk}

\begin{document}

\title{Lost in Algorithms}


\author{Andrew N. Sloss}
\affiliation{%
  \city{Seattle}
  \state{Washington State}
  \country{USA}
}
\email{andrew@sloss.net}




\begin{abstract}


Algorithms are becoming more capable, and with that comes \textit{hic sunt dracones} (\textit{\say{here be dragons}}). The term symbolizes areas beyond our known maps. We use this term since we are stepping into an exciting, potentially dangerous, and unknown area with algorithms. Our curiosity to understand the natural world drives our search for new methods. For this reason, it is crucial to explore this subject. \par

In this document, we look for future algorithms and styles. Each era in computing has had an algorithm focus. Examples include periods when military range prediction was necessary, \textit{weather prediction}, and, more recently, \textit{machine learning}. The 1940s saw the starting point when electronic machines replaced humans \cite{bhattacharya2022man}. Procedures became too complex to be handled by people. This time and other historical periods have accelerated additional specific algorithm development and the associated hardware architectures. The question for this paper is, \textit{what next?} As we explore this question, we will introduce a set of practical terms to help classify the various algorithms and hopefully provide a straightforward method of understanding. At the highest level, algorithms are recipes for solving problems. Problems range from the small \& simple to the large \& complex. \par

This project covers the behavior of algorithms and the hardware styles to execute those algorithms. In other words, we will cover \textit{what the algorithms do rather than how they do it}. The world of algorithms is a large and complicated subject. To assist in the process, we will separate the world into three main areas: \textit{computer science}, \textit{artificial intelligence}, and finally \textit{quantum computing}.\par

The project's objective is to overlay the information obtained, in conjunction with the state of hardware today, to see if we can determine the likely directions for future algorithms'. Even though we slightly cover non-classical computing in this paper, our primary focus is on classical computing (i.e., digital computers). It is worth noting that non-classical quantum computing requires classical computers to operate; they are not \textit{mutually exclusive}.

\end{abstract}


\settopmatter{printacmref=false}
\setcopyright{none}
\renewcommand\footnotetextcopyrightpermission[1]{}
\pagestyle{plain}


\begin{teaserfigure}
  \includegraphics[width=\textwidth]{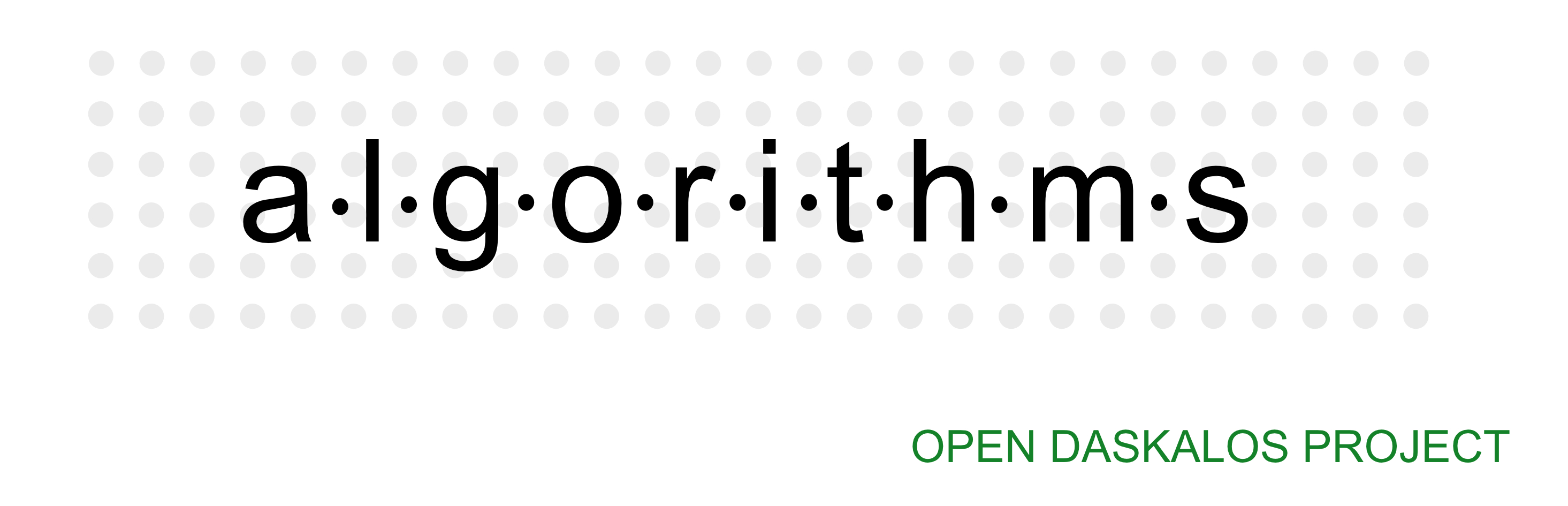}
  \label{fig:teaser}
\end{teaserfigure}

\maketitle

\section{Introduction}

\begin{figure}[ht!]
  \centering
  \includegraphics[width=0.8\linewidth]{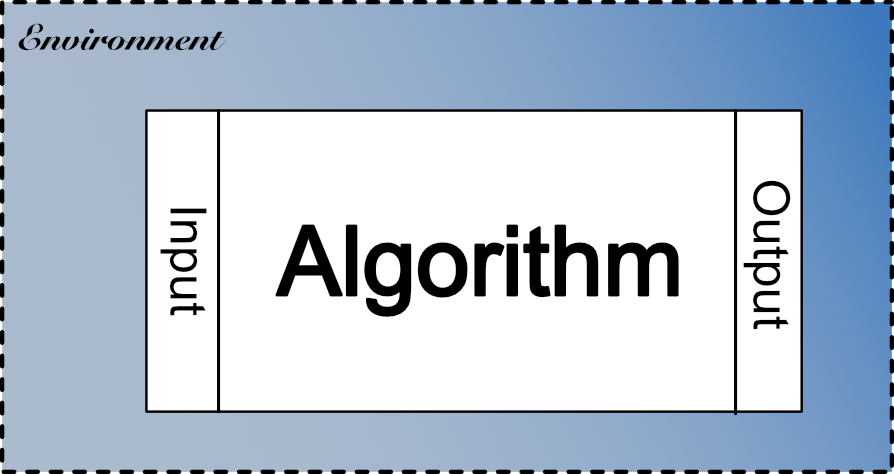}
\caption{Algorithm-Environment}
\label{figure:algoenv}
\end{figure}

Algorithms are critical for industry, research, and ideas. And have an increasing influence on society. We can say every part of our lives involves some form of an algorithm. It is a ubiquitous tool for problem-solving. To run an algorithm, we have many machines: mechanical, biological, analog, digital, quantum, and even human-social. We use them to optimize (e.g., repetition), explore (e.g., search), and even predict (e.g., models). The subject attracts some of the brightest and most innovative people wanting to discover better solutions. These people are always at the cutting edge, striving to do the next complicated task. \par

Figure \ref{figure:algoenv} shows how we will approach the subject. An algorithm or recipe has inputs and outputs and lives inside an environment. \par

The inputs-outputs come as part of the executing environment, e.g., a digital computer. We will look at algorithms from a behavioral viewpoint. In other words, \textit{what do the algorithms do?}, and less on \textit{how they do it?}. It is essential to understand the distinctions; we are not explaining the algorithms themselves but how they affect or take from the environment. Because of the vast nature of the field, the focus is mainly on digital computing to help prune the domain.\par

\begin{figure}[ht]
  \centering
  \includegraphics[width=0.8\linewidth]{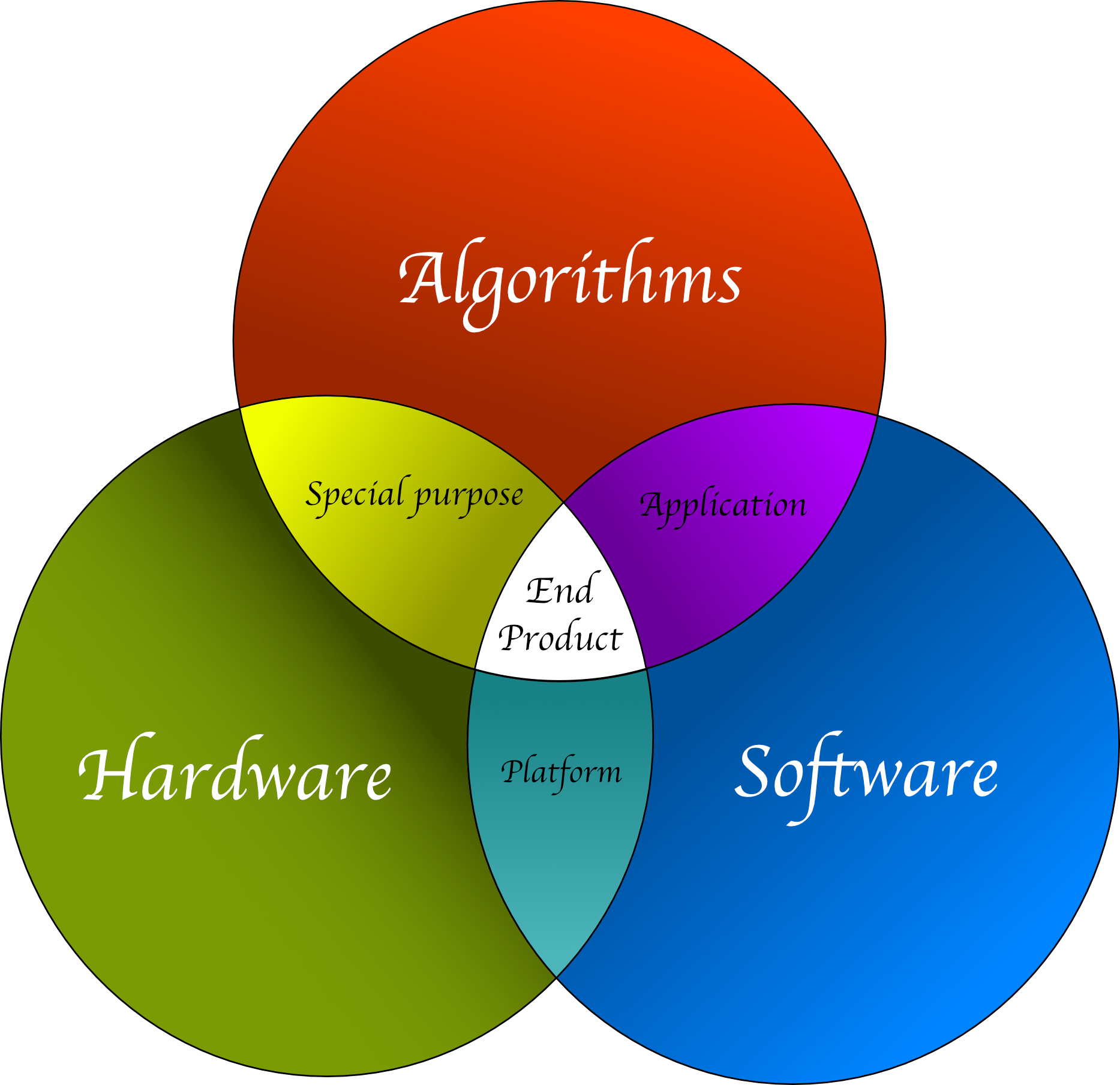}
\caption{Relationship between Hardware-Software-Algorithms}
\label{figure:relationship}
\end{figure}

We have hopefully explained our approach, and the next stage is to show the relationship between hardware and software. The Venn diagram shown in Figure \ref{figure:relationship} attempts to establish the relationship between hardware, software, and algorithms. We are separating the worlds, so it is easier to see the connections. Each world has its styles of thought and process. For instance, algorithms + software gives us applications, and hardware + software + algorithms give us an end product. A complete solution is about all three worlds aligning together. \par 

\textit{What came first, hardware or algorithms?} This is a \textit{chicken or egg} question. Algorithms are probably first, with the role of hardware being initially human and algorithms being mathematical equations. Historically, hardware has had to catch up with algorithms. This catch-up is why sizable effort is applied to optimize a new algorithm for existing hardware. With one exception, hardware jumps in capability every so often, forcing algorithms to catch up, e.g., quantum hardware. Hardware provides the enabling canvass for the exploration and optimization of algorithms.\par

An algorithm describes a method to utilize software and hardware for problem-solving. There is usually an \textit{objective}, but not always. A problem belongs to a \textit{problem domain}. A problem domain is a search space. For example, if a company has a problem with obtaining electronic components, the problem domain would probably include suppliers, the ordering process, and the manufacturing department. The algorithm should search those areas to determine a solution. The environment consists of the execution machine, data input, and a place for the final output decision. The execution machine is any computation system. Data input comes from the environment as filtered or noisy real-world data. Lastly, output decisions can be anything from classification to an action that makes an environmental change \par

As computer scientists, we have historically relied on famous texts. For instance, \textit{Robert Sedgewick et al.} book on \textit{Algorithms} \cite{Sedgewick2011}, or \textit{Donald Knuth} book series on \textit{The Art of Computer Programming} \cite{10.5555/260999} to provide libraries of known solutions. The knowledge includes such algorithms as \textit{recursive tree structure walking} or the \textit{shortest path} between two nodes on a graph. These libraries were the result of decades of experimentation. If we fast forward to today, we see \textit{statistical} and \textit{probabilistic} algorithms becoming ever more popular. These algorithms are less concerned with precision (i.e., absolutes) and more concerned with \textit{good enough} (i.e., levels-of-certainty). This trend does not mean the older algorithms are any less important, but currently, they are not at the cutting edge. \par

Mathematics allows us to express complex processes or prove correctness. It is probably our most significant accomplishment. The language of mathematics underpins the world of algorithms, and it is how we formally describe recipes. We will start this journey by looking first at the general objectives, i.e., \textit{what should algorithms explore?}.

\subsection{General objectives}\label{perfect}

\begin{figure*}[hbt!]
\begin{center}
\includegraphics[width=0.60\linewidth]{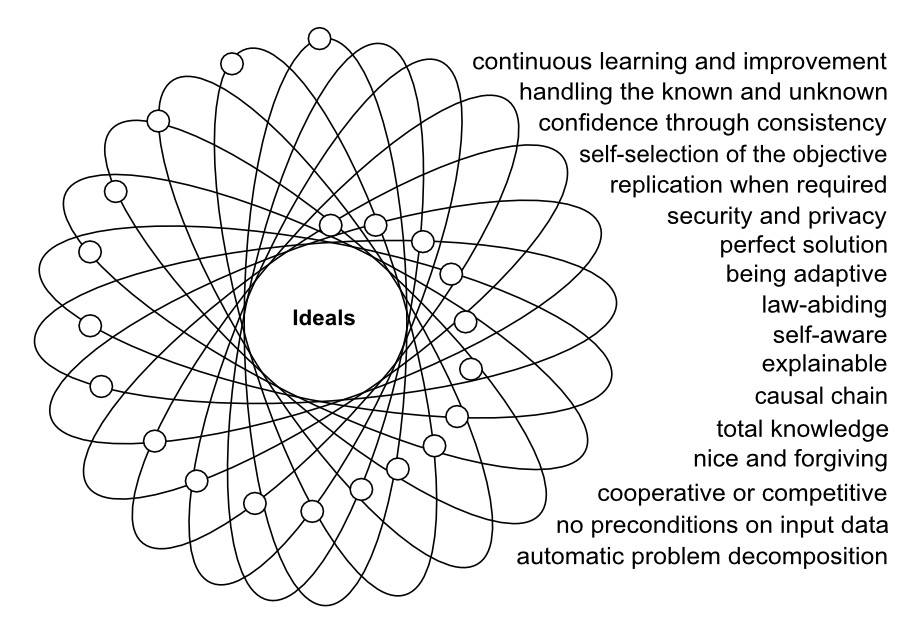}
\caption{Ideal algorithm}
\label{figure:perfect}
\end{center}
\end{figure*}

 Algorithms optimize, summarize and discover the world around us. These pursuits are related to us humans, either to augment or to move beyond our capabilities. We separate objectives into three distinct areas:\\
 
\begin{itemize}[leftmargin=*]
    \item \textbf{\textit{What we know}}
    
    Using historical knowledge and skills
    \item \textbf{\textit{What we think we want to know}}
    
    Using investigative methods of exploration
    \item \textbf{\textit{What we can't imagine}}
    
    Going beyond human perception and capability\\
\end{itemize}

In the beginning, we described an algorithm as a recipe. The recipe is a proven sequence of operations that find an answer or execute computation. Historically, algorithms were optimizations of what we knew \textemdash replicating human procedures. The early hardware had significant constraints, limiting what algorithms existed. These early solutions, even though primitive by today's standards, could operate 24 hours a day, seven days a week (provided the thermionic valves did not burn out). The hardware was a digital replacement. As time has progressed, improvements in computation have allowed us to shift from \say{\textit{what we know}} to \say{\textit{what we think we want to know}}. In other words, the hardware allows us to explore, i.e., find new knowledge. Exploring involves some form of guessing and, by implication, takes time. Guessing allows for mistakes. Lastly, this brings us to the third objective \say{\textit{what we can't imagine}}, these are the algorithms at the edge of a discovery that cannot necessarily be 100\% explained or show a reasoned causal path. Their forms and structures are still in flux.  \par

\textit{Certainty} requires some form of confirmation. We used the word \textit{proven}, in an earlier paragraph, as an aside. A mathematical proof is used for validation and differs from an algorithm. In that, what is computable and what can be proven may be different. We have an arsenal of automated mechanisms to help verify algorithms. These include formal methods and the modern trend to explore complexity hierarchies with various forms of computation. It is worth mentioning that \textit{Kurt Gödel}, in 1933, presented the infamous \textit{Incompleteness theorem}, showing that not all algorithms can be proven \cite{kennedy_2022}. Another critical example is \textit{David Hilbert's}'s \textit{Halting problem} \cite{turing1936a}, where \textit{Alan Turing} proved that it is impossible to determine when an algorithm will stop (or not) given an arbitrary program and inputs.\par

As the objectives become more abstract, \textit{uncertainty} increases. We can divide uncertainty into two ideas. There is uncertainty due to the complexity of nature, and there is uncertainty due to our lack of knowledge. Both play a critical role as we explore our environment. The first idea is called \textit{ontological uncertainty} (e.g., associated with biology and quantum mechanics), and the second idea is called \textit{epistemological uncertainty} (e.g., we do not know the precise number of people who are left-handed?) \cite{TIMP2022}. \par

\say{\textit{What we know}}, \say{\textit{what we think we want to know}}, and \say{\textit{what we can't imagine}} are the three high-level objectives; we next look at the aspirations. \textit{What should we consider as ideal attributes for a good algorithm?}

\subsection{Ideals}

There are many ways to think about algorithms. We can look at demands that an algorithm has to satisfy (e.g., best voice compression algorithm or highest security level for buying online). The approach we have decided to adopt is to look at the attributes we want algorithms to have, the set of ideals. \textit{Platonic idealism} is the contemplation of ideal forms. Or, more realistically, a subset of ideal forms. An ideal involves attempting to find an algorithm without sacrificing other essential attributes. These attributes include being efficient with time (the \textit{temporal} dimension) and using appropriate resources (the \textit{spatial} dimension). Resources include physical storage, communication, and computation. \par

We should make it clear this is not about what is possible with today's technology or even in the future but what we want algorithms to achieve, i.e., our expectations. The following list is our first attempt:

\begin{enumerate}[label={I-\arabic*.},leftmargin=*]
    \item \textbf{\textit{Perfect solution}}, is an obvious first ideal. A good outcome is to have several solutions, each providing a different path and varying levels of precision \& accuracy. Human biases, such as symmetry, are removed from the outcome unless there is a requirement for a human-biased result, i.e., a decision is made between impartial (ethical) or partial (practical) solutions \cite{Mozi}. Finally, the algorithm maps directly onto available hardware per the spatial dimension. 
    \item \textbf{Confidence through consistency}, we want consistency; an algorithm creates confidence by providing reliable results.
    \item \textbf{\textit{Self-selection of the objective}}, one of the essential activities humans undertake is determining the purpose. For an ideal algorithm, we want the goal or sub-goals set by the algorithm or offered as a set of options, i.e., negotiation.
    \item \textbf{\textit{Automatic problem decomposition}}, we want the algorithm to break down a problem into testable modules. The breakdown occurs automatically. This process is essential if we want to handle more significant issues, i.e., more extensive problems.  
    \item \textbf{\textit{Replication when required}}, specific problems lend themselves towards parallel processing. For these classes of problems, we want the algorithm to self-replicate. The replication allows solutions to scale automatically; some problems require scale. As much as possible, the algorithms should also work out how to scale linearly for a solution to be ideal. 
    \item \textbf{\textit{Handling the known and unknown}}, we want an algorithm to handle problems that are either known (with related solutions) or entirely unknown (where exploration occurs). An unknown answer, once found, transfers to the known. It learns.
    \item \textbf{\textit{No preconditions on input data}}, from an ideal perspective, we want to remove the format strictness imposed on the input data. Data acts as an interface for algorithm negotiations. Analyzing the data means the data format is deducible, removing the requirement of a fixed interface. Note Machine Learning has different criteria that are more to do with the quality of the input data.
   \item \textbf{\textit{Self-aware}}, ideally, we want an algorithm to be aware of the implications of a decision. This implication is especially true regarding safety-critical problems where a decision could have dire consequences and, more generally, the emotional or moral side of a decision. It weighs the effect of the outcome. 
    \item \textbf{Secure and private}, we want an algorithm to handle data so that it is secure, and if human information is concerned, it provides privacy. 
    \item \textbf{\textit{Being adaptive}}, an ideal algorithm can change as the environment changes and continuously learns from new knowledge. Knowledge comes from experimenting with the environment and subsequently improves the algorithm.
    \item \textbf{\textit{Causal chain}}, where applicable, we want the causal chain that produced the result. We want to understand \textit{why}.
    \item \textbf{\textit{Total knowledge}}, an ideal algorithm has all the necessary historical knowledge for a particular area. The algorithm does not follow information blindly but has all the knowledge about a specific subject. New knowledge can be identified as an emergent property if an unknown pattern appears. The ideal algorithm becomes an \textit{encyclopedia} on a particular subject. Maybe different weights are placed on the knowledge that is correct or contradictory. Maturity means the topic under study is wholly understood and has well-defined boundaries.
    \item \textbf{\textit{Explainable}}, we want results explained in human understandable terms. As the problem becomes more complicated, so do the answers. We want the algorithms to explain the answer and, potentially, the context.  
    \item \textbf{\textit{Continuous learning and improvement}}, as alluded to in previous ideals, we want the algorithm to continue to learn and to continue attempts to improve the techniques to produce a better, faster, less resource-draining solution.
    \item \textbf{\textit{Cooperative or competitive}}, the ideal algorithm works in a multi-agent environment, where agents are assistants for or detractors against a zero-sum scenario. \cite{10.5555/531517}—the ideal looks for an alliance with other agents. If an alliance is not possible, it goes out on its own to solve the problem (Nash equilibrium  \cite{9780262150415}). In other words, we want an algorithm to have \textit{Games Theory} skills.  
    \item \textbf{\textit{Law-abiding}}, we need an algorithm to be a law-abiding citizen. It works within the confines of legal law (not scientific laws). This confinement is essential for algorithms involved in safety-critical or financial activities.
    \item \textbf{\textit{Nice and forgiving}}, is more of a human constraint. We want algorithms to take the most friendly society approach in a multi-agent environment; if harm occurs to the algorithm, then a counter-reaction could be implemented. Within reason, an algorithm can forget any malicious act. This reaction is essential when only partial information is available \cite{10.5555/531517}. We can argue whether this is a constraint or an ideal, but we want algorithms to have some human-like tendencies, e.g., compassion over revenge). 
\end{enumerate}

As pointed out, these are first-pass ideals. We are certain ideals are missing or need modification. Even though algorithms are likely to be unique, ideal characteristics define the algorithms' boundaries (or extremes). The limits provide the universal goals for an algorithm.\par

We now move the journey to the problem domains. At this point, we have discussed the importance of algorithms and what the high-level ideals should be. Hopefully, these ideals indicate why we are potentially entering unknown territory. 

\subsection{Computation}\label{sec:computation}

Problems make an algorithm attractive, from the challenge of chasing a solution to the actual application. A solution is a map of a complex world, making it understandable. The map comes from a boundless library of ideas, e.g., \textit{The Library of Babel} concept \cite{borges2000library}. Each room in the infinite library includes varying truths and falsities. \par

As mentioned, problem domains determine the search space of possibilities, ranging from simple to complex and small to large. A simple mathematical problem domain tends to have a simple solution; likewise, a chaotic problem domain leans towards a complicated answer. There is an underlying belief and a hope that a problem domain is reducible \cite{Mitchell2009}. For example, \textit{Sir Isaac Newton} created the \textit{Laws of Motion} \cite{newton} that reduces the complexity of movement to a set of rules. By contrast, we have failed to reduce gravitation, electromagnetism, weak nuclear, and strong nuclear forces to an agreed-upon single solution, i.e., a \textit{Grand Unified Theory (GUT)}. We have controversial ideas, such as \textit{String Theory}, but no provable solutions \cite{kaku2021god}. There is a possibility and hope a future algorithm will eventually solve this problem. \par 

Algorithms rely on reducibility. The likelihood of finding a reducible form is dependent on the complexity level. How we reduce a problem also depends on the \textit{spatial} and \textit{temporal} constraints. A good example of an external constraint is the timing required for a successful commercial product. Missing the timing window means a potential loss of revenue. \par

\begin{figure}[ht]
\begin{center}
\includegraphics[width=0.95\linewidth]{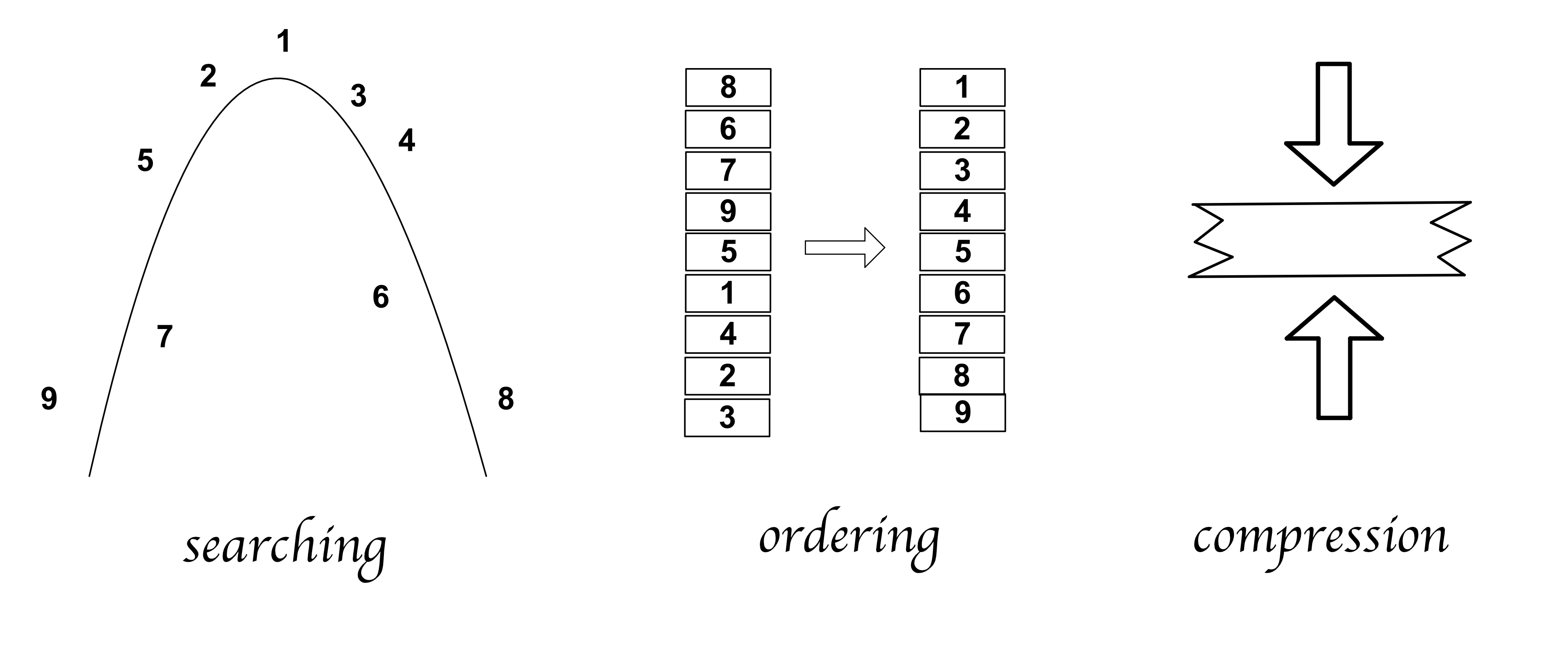}
\caption{Searching, ordering, and compression}
\label{figure:algo}
\end{center}
\end{figure}

A solution to a problem domain involves a combination, and sequence,  of \textit{searching}, \textit{ordering}, and \textit{compression}, see Figure \ref{figure:algo}. The searching function involves looking for a solution within the problem domain. Searching is about discovering \textit{knowledge}. The sorting function organizes the problem domain in a logical order, i.e., \textit{information}. Finally, the compression function converts the elements to a new form, i.e., \textit{meta-data}. These three functions find solutions to problems. \par

There are three \textit{embarrassing} worlds. We have gone over the goal of reducing complexity and the functions to map complexity to some form of reasoning; now, we look at the types of problem domains. We start with the first, \textit{embarrassingly parallel} problems; these are problems that map perfectly onto parallel solutions. These problem types do exist but are relatively rare. Performance is proportional to the available parallel machines, i.e., the more parallel machines available, the faster the processing. These improvements remain accurate, while the serial sections are minimal (i.e., taking into account \textit{Amdahl's law}, i.e., parallel performance becomes limited by the serial parts \cite{10.1145/1465482.1465560}). \par

The second embarrassing style is sequential data, i.e., \textit{embarrassingly sequential}. Embarrassingly sequential data follows a strict structure that can be mechanically optimized. We can design efficient computation to work best on embarrassingly sequential data. These problem spaces map efficiently onto software-hardware systems. They are less chaotic and random. \par

As we move to real-world situations, the domain types require complicated synchronizations (a mixture of parallel and serial components) and deal with unstructured data. The problems end up being \textit{embarrassingly unhelpful}. Bringing the focus back to hardware efficiency, the prior (i.e., embarrassingly parallel and sequential) leans towards \textit{specialization}. The latter (i.e., embarrassingly unhelpful) leans towards \textit{general-purpose} machines. General-purpose machines move towards the \textit{Principle of Universality}. As-in is capable of handling all problems. General-purpose machines are better for embarrassingly unhelpful problems since they reduce complexity using less specialized operations. The embarrassing aspects of data drive computation design. \par

\begin{figure}[ht!]
\begin{center}
\includegraphics[width=0.9\linewidth]{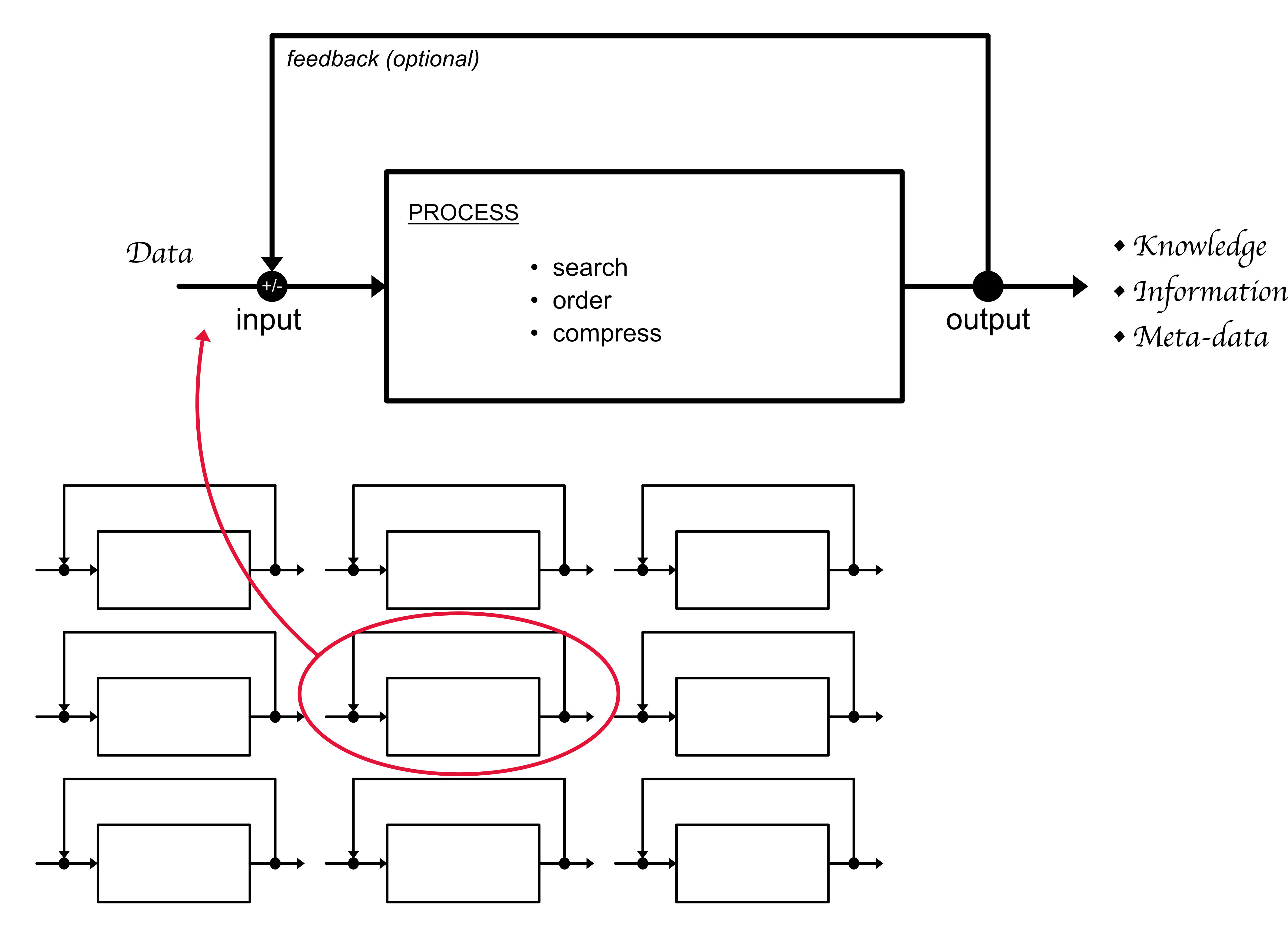}
\caption{Process function}
\label{figure:compsci}
\end{center}
\end{figure}

\textit{What is computation?} It is the act of running an algorithmic recipe on a machine. A computation process takes input data and outputs some form of result; see Figure \ref{figure:compsci}. A process can be serially sequenced or run in parallel. Optional auxiliary feedback is taken from the output and placed as input, giving an algorithm the ability to adapt. This characteristic allows for the creation of complex hardware architectures. Conditional control mechanisms (i.e., \textit{if-then-else}) determine the order and flow of the computation either between processes or within the process. 

\begin{figure}[ht] 
\begin{center}
\includegraphics[width=0.30\linewidth]{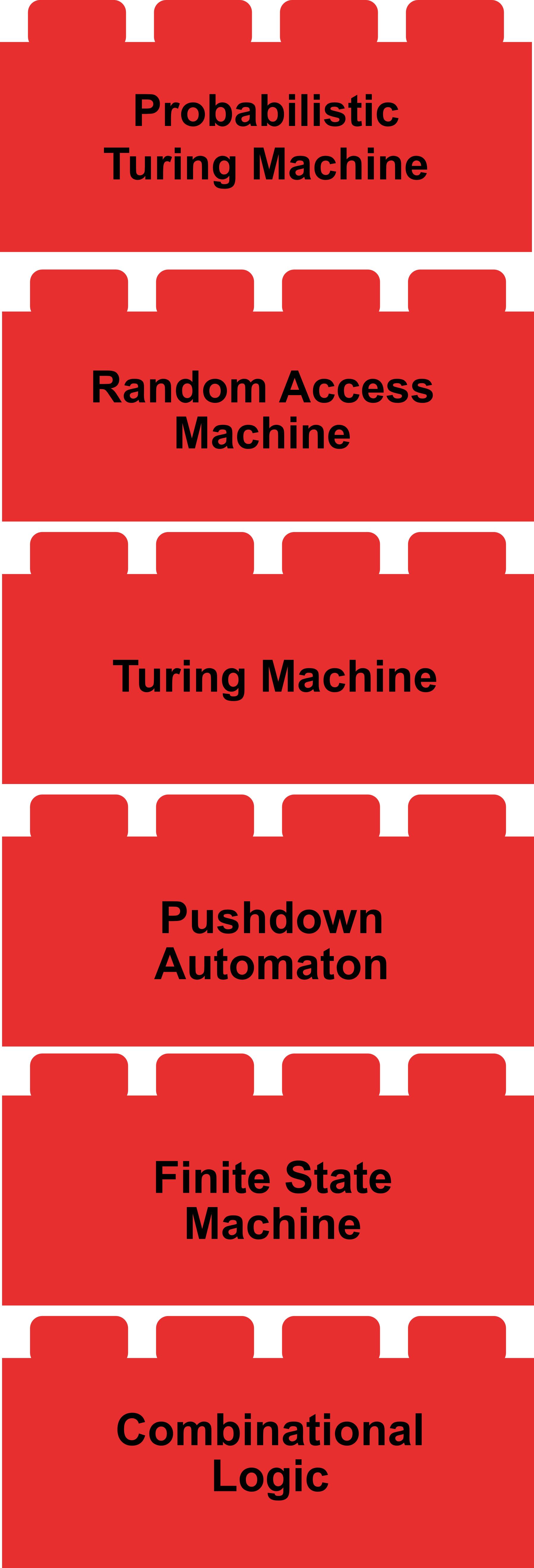}
\caption{Computation machines}
\label{figure:automon}
\end{center}
\end{figure}

Figure \ref{figure:automon} shows a subset of machines, with \textit{combinational logic} as a foundation for the other levels. In this diagram, we place \textit{Probabilistic Turing Machine} at the top of the computation stack. This decision will hopefully become self-apparent as we journey further into algorithms. \par

Energy is a requirement for computation and, as such, has direct influence on the choice of algorithms. Carrying out any calculation requires a form of energy imbalance (following the \textit{Laws of Thermodynamics}). To achieve energy imbalance in classical computing, we either supply energy directly or harvest the energy from the environment. Once energy is supplied, execution results in the production of heat. Maybe someday we can recycle that heat for further computation. With future machines, \textit{reversibility} may become an essential requirement to reduce energy and to allow more capable algorithms, i.e., \textit{reversible computing} \cite{LANDAUER2002}. \par

\subsection{Input-output relationship}

What are the general relationships between the input and output data? Below are some connections from simple to complex. Finding any relationship from the data, even a good-enough one, can be highly complicated. \\

\begin{itemize}[leftmargin=*]
    \item \textbf{\textit{Linear relationship}}, shows a straightforward relationship between input data and result, example equation $y=2x$.
    \item \textbf{\textit{Exponential relationship}}, shows a growth factor between input and output, example equation, $y=e^x$. 
    \item \textbf{\textit{Nonlinear relationship}}, difficult to determine the relationship, a more complex pattern is emerging, example equation $x^2+y^2= 42$
    \item \textbf{\textit{Chaotic relationship}}, at first sight appears to have no relationship, due to complexity, example equation $x_{t+1}=kx_t(1-x_t)$
    \item \textbf{\textit{Random/stochastic relationship}}, the relationship is truly random, example equation $y=rand(x)+42$ 
\end{itemize}

\subsection{Exploitative vs exploratory}

Algorithms handle the known or unknown, namely \textit{exploitative} or \textit{exploratory}. The first relies on knowledge and provides \textit{hopefully} a known outcome, for example, following an applied mathematics equation to determine whether a beam is in tension or compression. The second type, exploratory algorithms, explores a problem domain when the exact answer, and maybe even the environment, is unknown or changing \textemdash, for example, an algorithm learning to fly on a different planet. The alien world will have unknown gravitational or magnetic challenges. The former concept leans towards precision and accuracy, whereas the latter is comfortable with \textit{good-enough} results, i.e., compromise or palliative.\par

\begin{table*}[ht]
\begin{tabular}{ |l|l| } 
\hline
\textbf{Exploitative} & \textbf{Exploratory} \\
\hline
Specialized & General purpose \\
\hline
Narrow & Broad\\
\hline
Focused & Unfocused\\
\hline
Known, well understood & Unknown, and less understood\\

\hline
Best result & Good enough\\ 
\hline
Turing complete/incomplete & Turing complete\\ 
\hline
\end{tabular}
\caption{Exploitation vs Exploratory}
\label{table:comparison}
\end{table*}

\textit{What comes first, exploitative or exploratory?} This is a \textit{cart before the horse} question. Whether human or machine, exploratory takes place before exploitation. It is part of the learning process because we first have to understand before we can look for a solution. As we move into the future, we will rely more on algorithms to explore and find new solutions. And hence, hardware will need to increase support for more experimental methods. An example of increasing exploratory support is the trend towards efficient hardware for training \textit{learning} systems. Efficiencies in both speeds of training and power consumption. \par

Newer algorithms can take advantage of both techniques, i.e., explore first and then optimize or exploit until further exploration is required. This shift is a form of \textit{simulated annealing}. Annealing is the method of toughening metals using different cooling rates; simulation annealing is the algorithm equivalent. Forward simulated annealing starts by first exploring \textit{global} points (e.g. random jumps) and then shifts to \textit{local} points (e.g. simple movements) as the perceived solution becomes more visible. \par

\textbf{Hardware-software}: the software can play directly with exploitative and exploratory algorithms. By contrast, hardware is dedicated to the exploitative side, i.e., getting the most out of a known algorithm. Hardware is static and fixed. And the software provides the ability to adapt and re-configure dynamically. In the digital context, the software is the nearest we have to adaptive biological systems. Exploitation allows for hardware-software optimizations. \par

\subsection{Where do algorithms come from?}


\textit{Are algorithms entirely invented, or are they driven by the problems?} This is a time-old question with deep philosophical arguments from many sides. If we believe algorithms are invented, anticipating the future could be difficult. For this reason, we have chosen the view that problems define algorithms. To make this even easier, we will state that problems fall under the following four categories: \textit{i. Physics}, \textit{ii. Evolution}, \textit{iii. Biology}, and \textit{iv. Nature}.
Where algorithmic ideas develop from one or a combination of categories. \par

\begin{enumerate}[label={\roman*.},leftmargin=*]

\item \textbf{Physics} gives us the exploration of thermodynamics, quantum mechanics, and the fabric of the universe. Problems from a planetary scale to the sub-atomic 

\item \textbf{Evolution}, gives us exciting ways to create future options. Allows algorithms to explore their problem domains, i.e., survival of the fittest, natural selection, crossover, and mutation

\item \textbf{Biology}, introduces complex parallel networks, e.g., cell interactions and neuron communications

\item \textbf{Nature}, provides us with big system problems, e.g., climate change

\end{enumerate}

As previously pointed out, mathematics is the language to describe or express algorithms. It is not necessarily a vital source of inspiration. We believe that the discovery and understanding of the natural world create algorithms. By observing the natural world, we can play with predicting future directions and possibilities.

\subsection{Measuring computational complexity} \label{ss::computational}

\begin{figure}[ht]
\begin{center}
\includegraphics[width=0.8\linewidth]{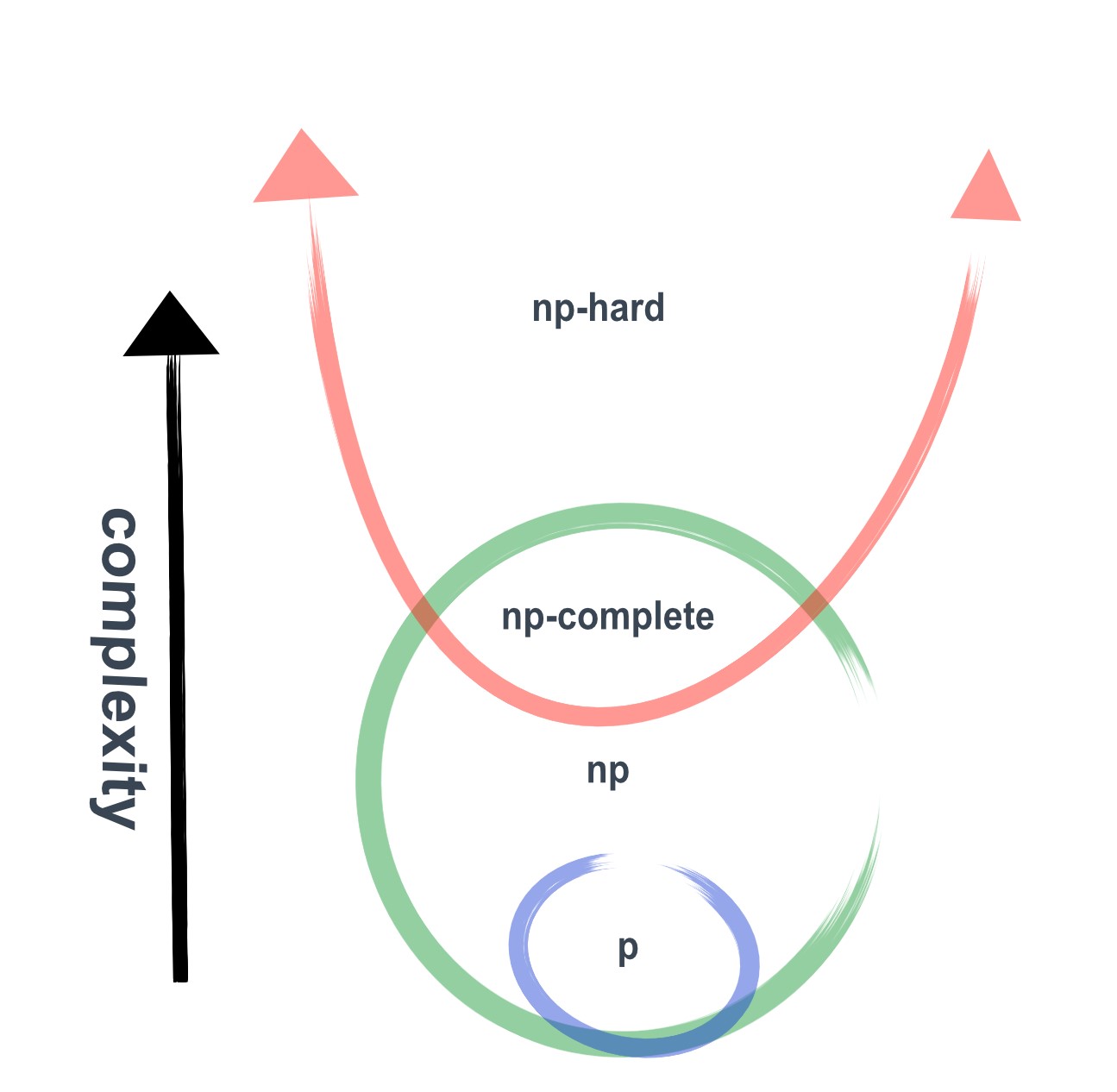}
\caption{A Euler diagram on complexity - Computer Science}
\label{figure:euler}
\end{center}
\end{figure}

Many subjects are concerned with complexity. For example, safety-critical systems are susceptible to increases in complexity. Computer science has an area of research labeled \textit{Computation Complexity Theory} dedicated to the subject. The theory translates complexity into \textit{time to solve}. It is worth mentioning that time taken and energy is closely connected. Our tentative goal is always to remain within an energy boundary. Problems break down into different time relationships as shown in the now infamous Euler diagram (see Figure \ref{figure:euler}). \textit{Why should we care?} Because there exist problems that are impossible to solve efficiently or are just unsolvable. Providing more engineering time or effort will not culminate in a faster solution in these cases. \par

\begin{itemize}[leftmargin=*]
\item \textbf{\textit{Polynomial} (P) time}, represents computational problems that are solvable in deterministic polynomial time. These problems are relatively straightforward to solve on a \textit{Turing machine} (see Section \ref{sec:computation}). Low in complexity. The input length determines the time required for an algorithm to produce a solution.

\item \textbf{\textit{Nondeterministic Polynomial } (NP) time}, are solvable problems but in nondeterministic polynomial-time. The algorithm can be proven correct using a deterministic Turing machine. Still, the search for the solution uses a nondeterministic Turing machine. The search involves some form of best guess.

\item \textbf{\textit{Nondeterministic Polynomial-complete}  (NP-complete) time}, similar to NP problem, the verification can occur in quick polynomial time but solution requires a \textit{brute-force} algorithm. Brute force means there is no efficient path to a solution. These problems are the most complicated to solve in the NP set. Also, each problem is reducible in polynomial time. The reducibility allows for simulation.  

\item \textbf{\textit{Nondeterministic Polynomial-Hard} (NP-hard) time}, covers the truly difficult problems, the hardest NP problems and continues outside the NP scope. It also includes potential problems that may not have an answer.
\end{itemize}

Suppose we look at complexity through an exploitative and exploratory lens. We see that \textbf{P} covers the exploitative algorithms, and generally, \textbf{NP} covers the exploratory side, i.e., experimentation. 

\subsection{Measuring probabilistic complexity}\label{probcomplex}

Continuing our journey, it becomes apparent that probability is becoming increasingly important. For this reason, we should try to understand probabilistic complexity, where \textit{good enough}, \textit{averages}, and \textit{certainty} play a more critical role. Probability is fundamental for artificial intelligence, probabilistic computers \cite{PROBCOMPUTER}, and quantum computing, as described later. \par

These are problems only solvable by a \textit{probabilistic Turing machine}  \cite{Probabilistic}. This Turing machine operates with a \textit{probability distribution}, which means that a distribution governs the transitions. This setup gives a probabilistic Turing machine-specific unique characteristics. \par

The characteristics are nondeterministic behavior, transition availability by a probability distribution, and stochastic (random) results. The stochastic effects require repeatability before a level of certainty can be established. The behavior means that the same input and algorithm may produce different run times. There is also a potential that the machine will fail to halt. Or the inputs are accepted or rejected on the same machine with varying execution runs. This variability is why an average of execution runs is required. \par

\begin{figure}[ht]
\begin{center}
\includegraphics[width=0.9\linewidth]{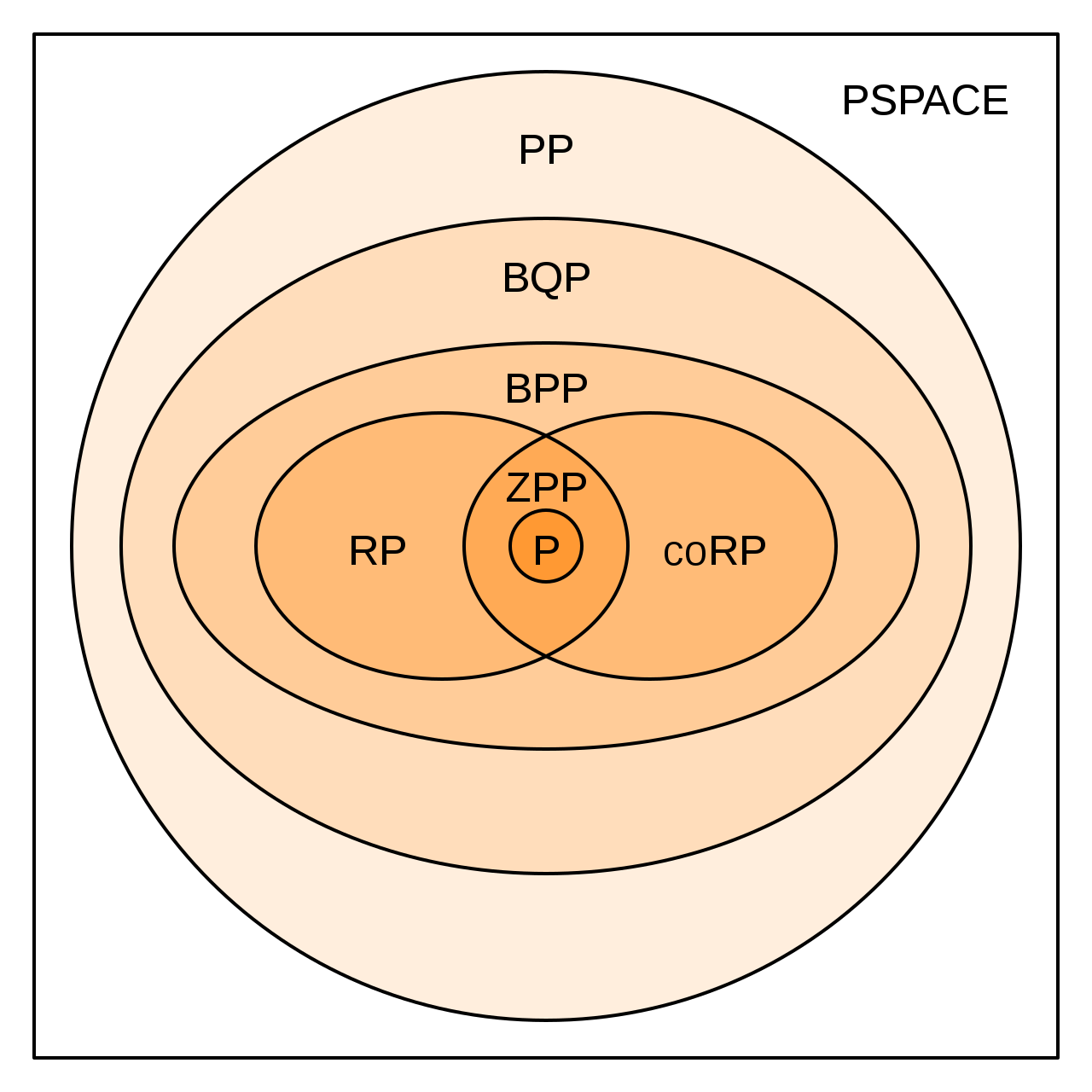}
\caption{Probabilistic complexity (Wikipedia)}
\label{figure:probablistic}
\end{center}
\end{figure}

Figure \ref{figure:probablistic} shows an extension to the traditional Euler diagram with probabilistic complexity  \cite{Probabilistic}. Rather than covering all the different options, permit us to focus on three especially relevant examples: 

\begin{itemize}[leftmargin=*]
  
\item \textbf{\textit{Bounded-error Probabilistic Polynomial} (BPP) time}, 
an algorithm that is a member of BPP runs on a classical computer to make arbitrary decisions that run in polynomial time. The probability of an answer being wrong is at most \( \frac{1}{3} \), whether the answer is heads or tails (i.e., a coin flip).

\item \textbf{\textit{Pounded-error quantum Polynomial} (BQP) time},

an algorithm that is a member of BQP can decide if there exists a quantum algorithm running on a quantum computer with high probability and guaranteed to run in polynomial time. As with BPP, the algorithm's run will correctly solve the decision problem with a probability of at least \( \frac{2}{3} \).

\item \textbf{\textit{Probabilistic Turing machine in Polynomial time} (PP) time},
is simply a algorithm were the probability of error is less than \( \frac{1}{2} \) for all instances.
\end{itemize}

Finally, probability complexity is linked directly to the probability of being wrong. Errors are part of the process and, as such, need to be handled or mitigated.  \par

Note that Figure \ref{figure:probablistic} captures the relative topology, but not the algorithmic class size. How important each member will be in comparison is still to be determined.\par

Now that we have described some parts of measuring complexity, we can move on to the objective.

\subsection{Algorithm objective}\label{ss:objective}

\begin{figure}[ht]
\begin{center}
\includegraphics[width=0.95\linewidth]{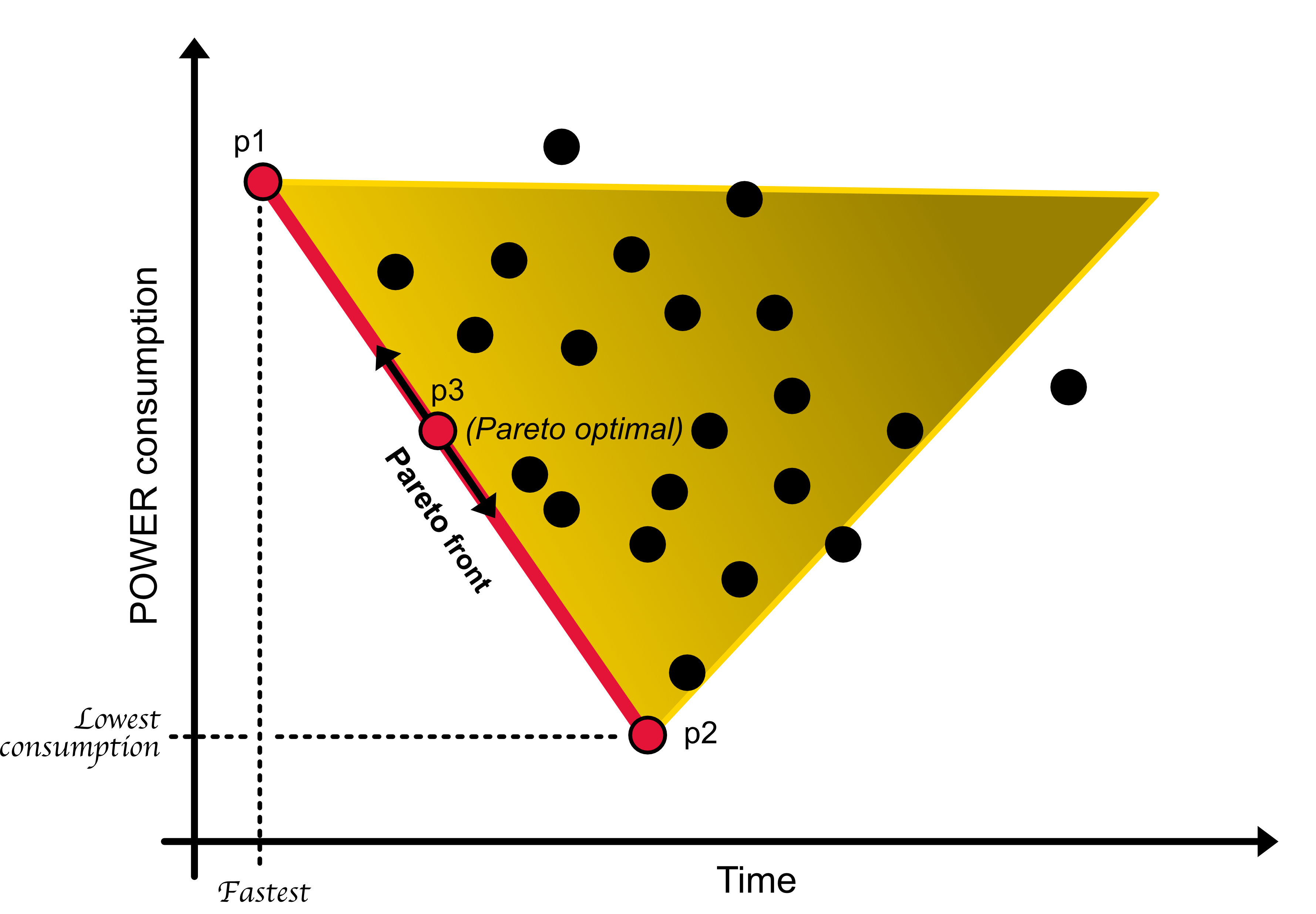}
\caption{Multi-objective Pareto Curve}
\label{figure:pareto}
\end{center}
\end{figure}

An algorithm has to have some form of direction. The direction can take one of three forms \textit{single objective}, \textit{multi-objective}, and \textit{objective-less}. A single objective means only one search goal, e.g., performance or power. Single search goals tend to be simpler problems to solve but not always. A multi-objective is more complicated with multiple search goals to satisfy, e.g., performance, power, and area. Multi-objective searches look for a compromise between the solutions. These compromises live on what is called the \textit{Pareto optimal}, see Figure \ref{figure:pareto}. Finally, objective-less relies on gaining new experiences in an environment rather than moving towards any particular goal \cite{10.5555/2792412}. The idea is that by gaining experience, a better understanding of the problem domain occurs, thus allowing for significantly better solutions.  \par

For most of this century, the focus has been on a single objective, but problems have changed in the last fifty years, and multi-objective problems are more typical. Figure \ref{figure:pareto} shows two objectives for an electric car. These objectives are the best acceleration on the x-axis and the lowest power consumption on the y-axis. \textbf{p1} represents the fastest option (e.g., fast tires, higher voltage, and performance electric motors), whereas \textbf{p2} represents the lowest power consumption (e.g., aerodynamic tires, voltage limited, and balanced electric motors). The edge between the two extremes is the \textit{Pareto front}, a point on the front (e.g., \textbf{p3}) is \textit{Pareto optimal}. The best solution must be a compromise between acceleration times and power consumption. There are no perfect solutions, just compromises of opposing objectives. \par

As a counter-intuitive idea, objective-less is an exciting alternative. The algorithm is placed in an environment and then left to discover. Progress occurs when a new experience is discovered and recorded. An example of this exploration style is \textit{Novelty Search} (see Section \ref{ai:novelty}). Potentially these algorithms can find new knowledge. The objective-based algorithms look for something known, and the solution is biased by implication. Whereas objective-less algorithms learn by experience, removing implicit bias.\par

As well as the objectives, there are constraints (or limitations). These constraints impose restrictions on any possible solution. For example, the conditions may include limited resource availability, specific time windows, or simply restrictions on power consumption. A general mathematical goal is to provide \textit{constraint satisfaction}, where each object in the potential solution must satisfy the constraints.

\subsection{Environment and data}

\begin{figure}[ht]
\begin{center}
\includegraphics[width=1.0\linewidth]{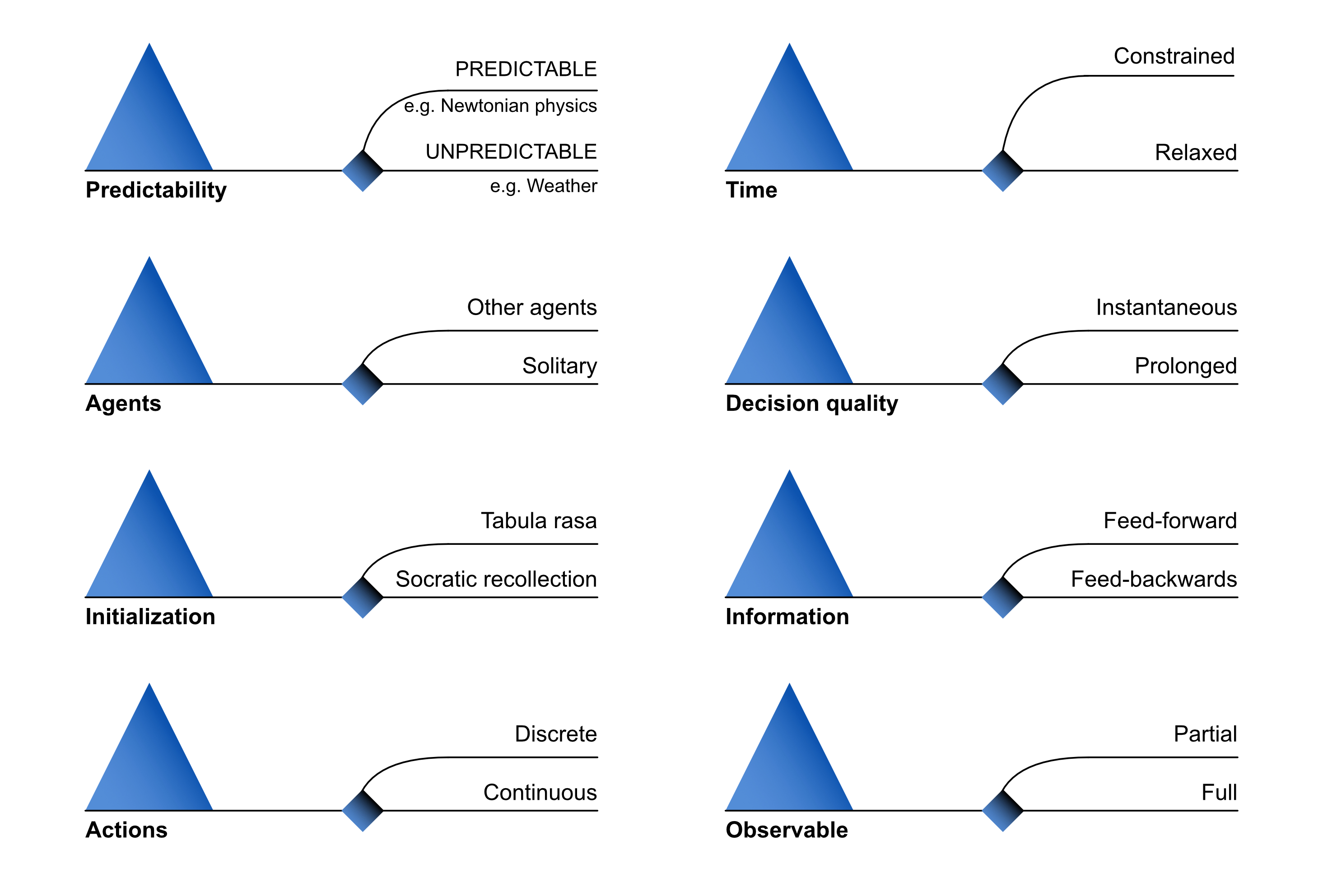}
\caption{Landscape \cite{russell2019human}}
\label{figure:landscape}
\end{center}
\end{figure}

There are many properties concerned with problem-domain landscapes. The first is the environment the algorithms operate within; see Figure \ref{figure:landscape}. For example, one environment could have the properties of solitary, relaxed, and completely observable. A domain can make the problem space more or less difficult to cover. The landscape diagram shows some of the variations. The variations act as a filter to determine which class of algorithms is more likely to be successful and rule out other ones that are unlikely. It seems common sense that an algorithm \textit{should} be chosen by first assessing the environment.

\begin{figure}[ht]
\begin{center}
\includegraphics[width=0.95\linewidth]{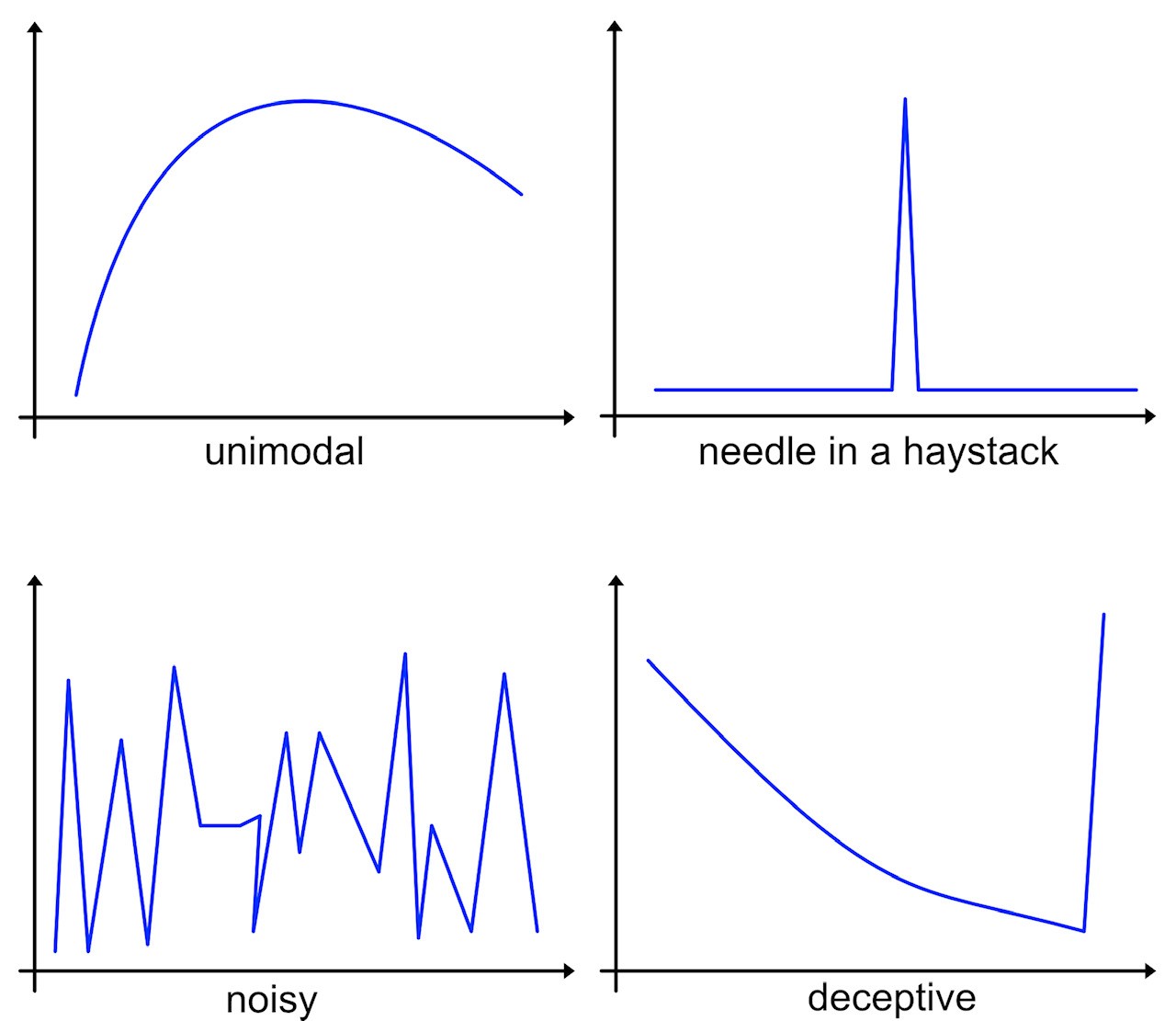}
\caption{Problem style \cite{Luke2013Metaheuristics}}
\label{figure:landscapeprob}
\end{center}
\end{figure}

Data comes in many different forms. From simple \textit{unimodal} data with an obvious solution to the more complex data that includes noise or \textit{deception} \cite{Luke2013Metaheuristics}, see Figure \ref{figure:landscapeprob}. The format of the input data determines the algorithm.  

By increasing the dimensions of the input data, we can extract hidden information \cite{reservoir}. More advanced algorithms use this technique to handle more complicated problems. The hope is to remove some of the noise in the data. The opposite approach can also be valid; reducing dimensions simplifies the input data.

\subsection{Determinism, repeatability, and randomness}

Algorithms have different characteristics, these include \textit{determinism}, \textit{repeatability}, and \textit{randomness}. Taking each characteristic in turn. Determinism is when an algorithm given a set of inputs provides a result in the same amount of time. Consistency is essential for applications that are time constrained. These applications come under the term \textit{Real-Time}. Where the algorithm flow consistently takes the same amount of time to complete. Real-Time is an arbitrary measurement since the definition of time can cover a wide range. \par

Repeatability is the concept that given the same input, the same output occurs. Many problems require this type of characteristic. It is fundamental to most mathematical equations. It is more aligned with perfection and means that the problem space is wholly understood. \par

Randomness is the opposite of repeatability, as the results can differ or even not occur. Randomness allows some degree of uncertainty to provide variation in the answers, i.e., flexibility in discovery. It goes against mathematical perfection because it allows for greater exploration of complex spaces. Algorithmic complexity can appear random because the patterns are so difficult to comprehend.\par

Random numbers in computers are called \textit{pseudo-random} numbers. This label is because they follow some form of artificial distribution. A \textit{pseudo random number generator} (PRNG) creates these numbers. As an important example, pseudo-random numbers can simulate noise. The simulated noise can help transition a high-dimensional problem into a more accessible lower-dimensional problem \cite{TIMP2022}. Achieving this transition occurs by replacing some of the state variables with guesses. This transition makes an otherwise impossible situation searchable (e.g., weather prediction).

\subsection{Prediction, causality, and counterfactual}

\textit{Prediction}, \textit{causality}, and \textit{counterfactual} are at the cutting edge of what algorithms are capable of achieving. Prediction is probably one of the most exciting areas for modern algorithms—the ability to predict a future with some degree of certainty. Science as a discipline has had its challenges with prediction. Prediction is a difficult subject; it includes everything from software to determine the next actions for a self-driving car to consistent economic forecasting on a specific stock. Probably the best-known of all the algorithmic predictions is weather prediction. Weather prediction is highly accurate for the next three hours but becomes less certain as we increase the time. \par

Casualty is the ability to show cause-and-effect \cite{Pearl}. The reason a pencil moved was that a person pushed the pencil. In many ways, humans want more than just an answer from an algorithm; they want to understand why. It is problematic for algorithms because it requires more understanding of how actions are connected and chained together. \par

Finally, there are counterfactuals—a combination of prediction and causality. Counterfactual is an alternative history where a decision not to do something affects a prognosis. This action can play with the future. Again an exciting area to play with from an algorithm point of view \cite{Pearl}. 
\subsection{Creativity and diversity}\label{sec:creativity}

Creativity and diversity are terms primarily associated with humans rather than algorithms. Creativity is some form of inspirational jump that allows complex problems to be solved. There is an ongoing debate whether an algorithm can be creative \cite{AIArt}? And if so, how do we measure creativity? If an algorithm paints a scenery, is it being creative? These are difficult questions, but when it comes to algorithms, this is the new frontier. 

\begin{figure}[ht]
\begin{center}
\includegraphics[width=0.9\linewidth]{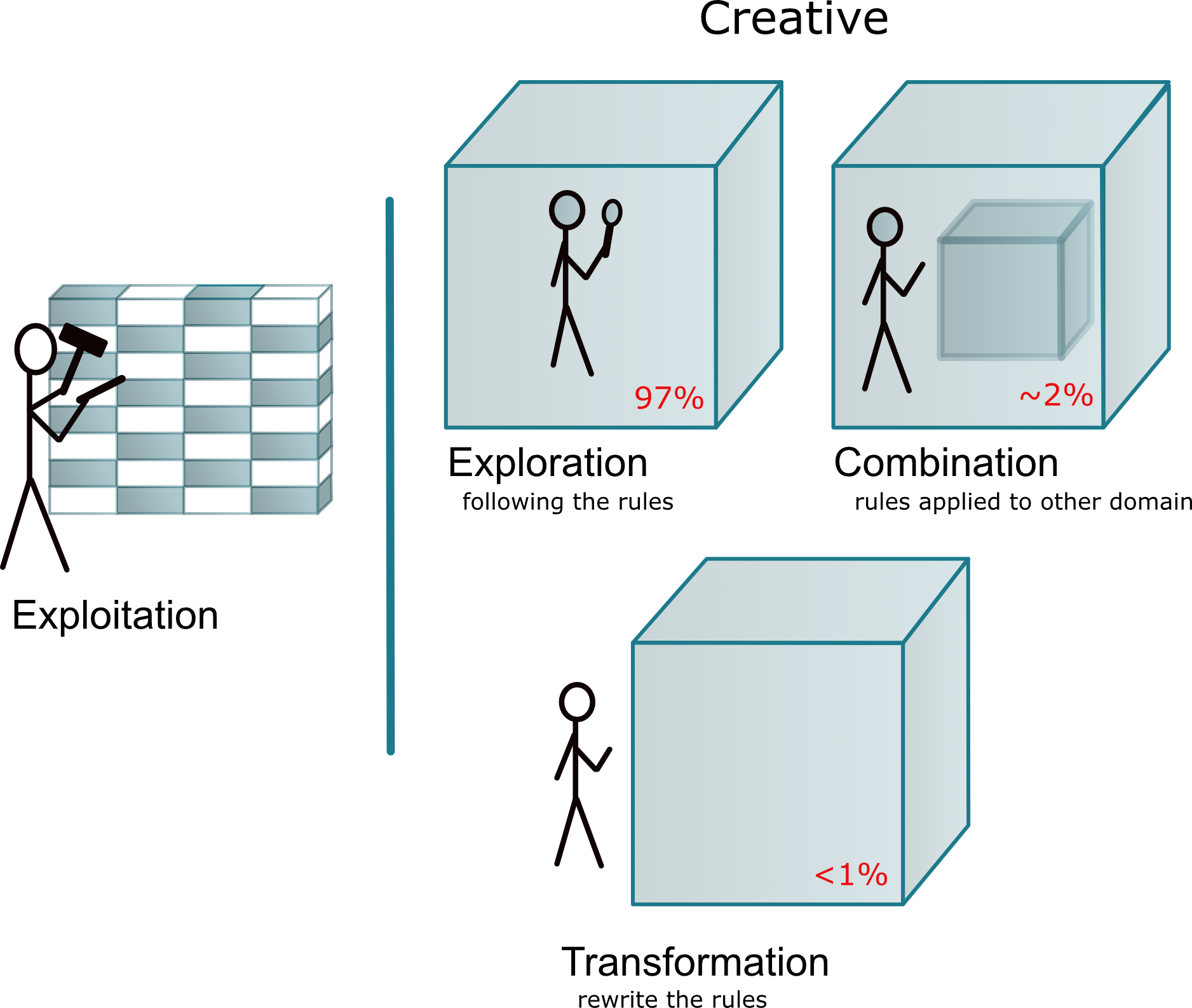}
\caption{Creativity \cite{Boden2013}}
\label{figure:creativity}
\end{center}
\end{figure}

What is creativity? \textit{Margaret Boden}, a Research Professor of Cognitive Science, broke down creativity into three useful mechanisms \cite{Boden2013}, namely \textit{exploration} - playing within the rules, \textit{combination} - applying one set of rules to another domain, and \textit{transformation} - rewriting the rules by removing a critical constraint. See Figure \ref{figure:creativity}. For algorithms, exploration creativity is risk-averse and limited, and at the other end of the scale, the transformation has the highest risk with potential novelty.\par

On the same lines of creativity, we have diversity. Diversity brings about novelty or new solutions by offering variation in the algorithms. A diverse set of algorithms can search multiple directions in parallel. \par
\subsection{Byzantine style algorithms}

Distributed systems, safety-critical systems, and financial systems need to have some resilience. Resilience is an attempt to avoid wrong decisions. Wrong decisions occur due to system errors or an intentional lousy agent. This area is called the \textit{Byzantine General Problem}, description outlined below:\par

\begin{enumerate}[label={B-\arabic*.},leftmargin=*]
    \item \textit{Lieutenant generals need to come to a decision.}
    \item \textit{Unfortunately, there are potential traitors.}
    \item \textit{How do the loyal generals decide on a correct decision?} 
\end{enumerate}

Redundancy is essential in several areas. For example, hardware or software has the potential to exhibit problems. It is necessary if the ramifications of such a situation occur to be able to mitigate those problems. This avoidance is crucial in safety-critical systems where failure results in harm or significant financial loss. The solution is to have redundancy and not rely on one or two agents for a decision. 
\subsection{No free lunch theorem}\label{nofreelunch}

When we described the Ideals in Section \ref{perfect}, we were skirting around the concept of a \textit{free lunch}. This idealism is a reverse play on the \textit{No Free Lunch} (NFL) theorem. David Wolpert and William McReady formalized the theorem, and it states that \say{\textit{all optimization algorithms perform equally well when averaged over all possible problems}} \cite{WolpMacr97}. The theorem means no algorithms stand out as being better or worse than any other. Solving different problems involves specific knowledge of that problem area.\par  

A modern algorithm has to deal with a knowledge question. The question is whether an algorithm starts from a clean slate (i.e., no knowledge \textemdash \textit{Tabula rasa}) or some known captured experience (i.e., known knowledge \textemdash \textit{Socratic recollection}). We need to decide how an algorithm starts; by learning with no expertise or giving the algorithm, a jump starts with knowledge. \par

\subsection{Network thinking}\label{sec:network}

In algorithms, it is crucial to mention networks. Networks play an essential role in modeling and analyzing complex systems. Networks are everywhere, from the interconnection between neurons in the brain to aircraft flight patterns between airports. Electrical engineering uses networks to design circuitry. Probably one of the most famous technology networks is \textit{The internet} which allows us to communicate efficiently. We commonly describe networks in terms of \textit{nodes} and \textit{links}. There are at least two helpful methods of describing networks: \par 

\begin{itemize}[leftmargin=*]

 \item \textbf{Small-world networks} are networks where all nodes are closely connected, i.e., requiring only a small number of jumps. Most nodes are not neighbors but are closely linked. For example, the aircraft flight patterns we already mentioned. Another example is Karinthy's 1929 concept of \textit{six degrees of separation} \cite{frigya1929,TIMP2022}, where just six links connect everyone. 
 
 \item \textbf{A scale-free network}, follows a \textit{Power law} distribution. The Power law states that a change in one quantity will cause a proportional change in another. All changes are relative—for example, a social network.

\end{itemize}

\textit{Preferential attachment} is a process where a quantity distributes across several nodes. Where each node already has value. The nodes with more value gain even more, and the nodes with less value gain less. This value transfer is necessary when algorithms model \textit{wealth distribution} or \textit{contribution-rewards in an organization}.\par

Lastly, network thinking is all about \textit{modern graph theory}. Graph theory covers \textit{graph knowledge models} to \textit{graph databases} (e.g., temporal-spacial databases). It is an important area for algorithms, and it is constantly expanding. 
\section{Category}

In this section, we will attempt to describe algorithms as categories or classes; it is a complicated process. Again more from the behavioral viewpoint. We will divide the world into three main categories \textit{computer science}, \textit{artificial intelligence}, and \textit{quantum computing}. We cover what we think are the more interesting behavioral classes, but this is by no means exhaustive. 


\subsection{Computer science, Cs}

Computer science has been creating and formulating algorithms for the past 50 years. This length of time means there is an abundance of algorithms. In this subsection, we collect the various essential concepts. As an academic subject, computer science is still relatively young as a discipline, but it acts as a universal provider. As-in it provides a service to all other fields. Note we included the first two descriptions as fundamental concepts rather than algorithms.

\subsubsection{Cs, Propositional logic}

Proposition logic is a language. It is used by algorithms to provide  \textit{Boolean} answers, i.e., \textit{True} or \textit{False}. Combining logic (\textit{OR}, \textit{AND}, \textit{Exclusive OR}, and \textit{NOT} operations) together we can create an algorithm. Digital hardware circuits derive from propositional logic.\par

\subsubsection{Cs, Predicate calculus} 

Predicate calculus is a language. Algorithms use it to produce correct statements. This correctness means that all statements are provable and true within the algorithm. Symbols represent logical statements. For example, $\forall x \in N :x^2 \geq x$ translates to \say{\textit{for all $x$, where $x$ is a natural number, it is true that $x^2$ is equal or greater than $x$}}. Thus satisfying the predicate calculus rules that all statements are sound and true. Another example, $\exists x \in N :x \ge 42$ translates to \say{\textit{there exists an $x$, where $x$ is a natural number, that $x$ is greater than $42$}}, again sound and true.

\subsubsection{Cs, Recursive algorithms}

A \textit{recursive algorithm} calls itself. For example, $f(x)=x-f(x-1)$, where the function $f$ is on both sides of the equation. Recursive algorithms have interesting behavioral properties because they can converge or diverge. A convergent recursive function concludes. A divergent recursive algorithm never stops.
In computer terms, memory resources can be pre-calculated in a convergent algorithm. Divergence means the opposite; an equation will fail to conclude, potentially resulting in an \textit{out of memory} error. We can use \textit{proof-by-induction} to determine correctness. \par

\begin{figure}[ht]
\begin{center}
\includegraphics[width=0.9\linewidth]{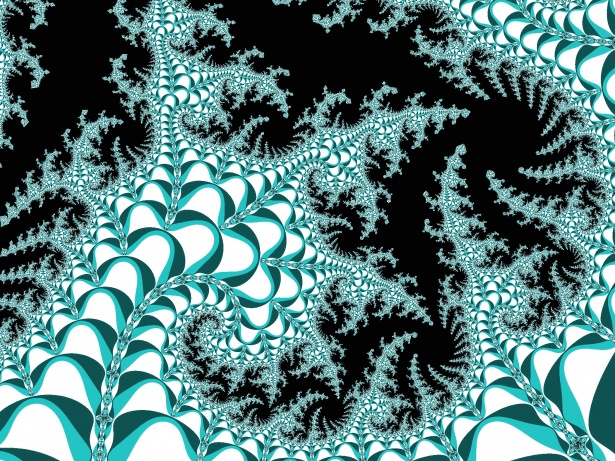}
\caption{Recursive fractal (Public Domain)}
\label{figure:fractal}
\end{center}
\end{figure}

Recursion can be seen in many natural objects, for instance, leaves, trees, and snowflakes. These occurrences mean there are many examples in nature where recursion is employed. Figure \ref{figure:fractal} shows recursion in the form of a fractal.\par

\subsubsection{Cs, Divide-and-conquer algorithms} \label{divide-and-conquer}

The \textit{divide-and-conquer} algorithms separate a problem into easier-to-manage problems. They help handle situations that are either too big in their entirety or too difficult.  

\begin{figure}[ht]
\begin{center}
\includegraphics[width=0.9\linewidth]{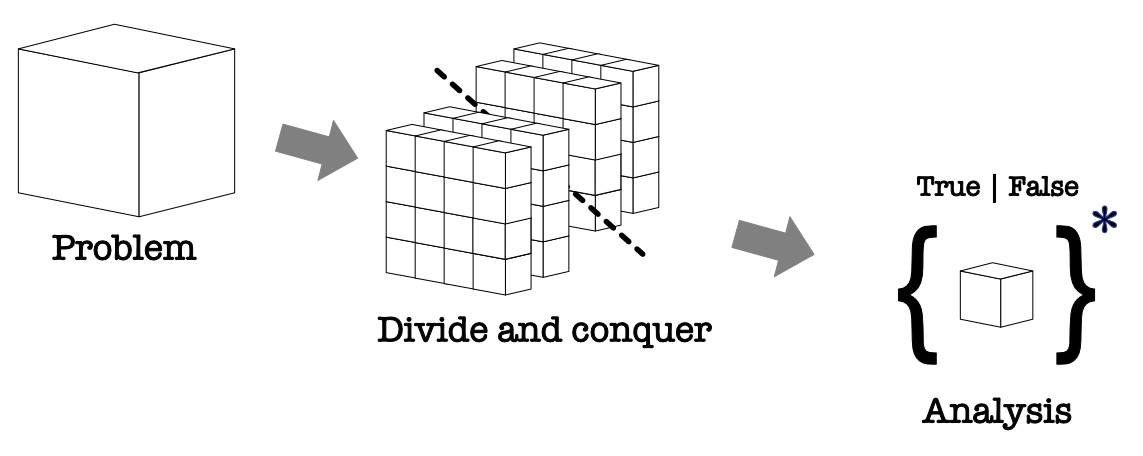}
\caption{Divide and conquer}
\label{figure:conquer}
\end{center}
\end{figure}

These algorithms are best employed when the divided problem is less complicated than the added communication and distribution overhead. If true, a parallel architecture can lend itself to this type of problem. And this is especially true if the sub-problems are all solved deterministically, i.e., taking the same time to process. Divide-and-conquer serves as a helpful method for debugging complex systems. This technique is closely related to the \textit{Scientific Method}.

\subsubsection{Cs, Dynamic programming algorithms}

\textit{Dynamic programming} (or DP), is related to the divide-and-conquer algorithms, see Section \ref{divide-and-conquer}. Unlike the divide-and-conquer, DP uses the result of the sub-problems to find the optimum solution to the main problem. These algorithms find the shortest path (e.g., map navigation), optimal search solutions, and operating system schedulers. 

\subsubsection{Cs, Randomized algorithm} Randomized algorithms use artificially generated randomness to solve complex problems, e.g., molecular interactions. One of the most famous algorithms in this class is called the \textit{Monte Carlo} algorithm. Monte Carlo takes an initial configuration, let us call it the \textit{status quo}, and, driven by a probability distribution, a state is randomly changed (e.g., on a coin, heads go to tails) \cite{miller2016crude}. A calculation then determines the energy of the new configuration. With that information, the energy acceptance criteria can determine whether the new configuration becomes the status quo. These algorithms model \textit{probabilistic real-world} systems. And as such require significant amounts of computational resources, not to mention the time is taken to set them up. Their primary use is in particle Physics, Biochemistry, and financial modeling.

\subsubsection{Cs, Fractional factorial design}

\textit{Fractional factorial designs} are included primarily because of their behavioral characteristics. They use a concept called \textit{sparsity-of-effects} principle \cite{surhone2011sparsity}. This principle brings out important features from the data. The claim is that only a fraction of the processing extracts the most interesting data features. This characteristic is an important behavioral style for many problem domains, i.e., using less work to identify the most interesting features. 

\subsubsection{Cs, Greedy algorithms}

A \textit{greedy algorithms} discover a solution a little piece at a time. These algorithms find near-optimal solutions for NP-Hard style problems. The algorithm takes the direction of the most advantage at each point (i.e., local optima strategy). This algorithm is relatively easy to implement for optimization problems.

\subsubsection{Cs, Brute force algorithm}

A \textit{brute force} algorithm, as the name implies, involves looking at every possible solution to find the best. In other words, no shortcuts. We use these algorithms when there are no alternatives. From a behavioral point of view, running these algorithms requires significant resources. These algorithms best suit problems with no known better solution and can get away with no time constraints. We default to these algorithms when there is no other solution. Replacement of these algorithms occurs when there are economic or environmental pressures. For example, we can see this with \textit{cryptocurrencies} as they change from brute force to be more environmentally friendly methods, i.e., \textit{proof-of-work} to \textit{proof-of-stake} \cite{proofofstake}.

\subsubsection{Cs, Backtracking algorithm}

\textit{Backtracking} algorithms move forward step by step. Constraints control each step. If a potential solution cannot achieve its objective, the path halts, and the algorithm backtracks to explore another possible solution path. This approach is robust at exploring different options for success. This algorithm can stop early if a solution option reaches a good-enough level of success.

\subsubsection{Cs, Graph traversal algorithms}

\textit{Graph traversal} is simply the process of visiting nodes in a graph. The visit can involve either reading or updating the node. There are different methods on the order of visits, e.g., \textit{depth-first} and \textit{breadth-first}. Where depth-first attempts systematically to visit the farthest nodes, by contrast, breadth-first attempts to visit all the nearest nodes first. Graphs are popular for many applications, including graph databases, spatial graph databases, and spatial-temporal graph databases.

\subsubsection{Cs, Shortest path algorithm}

Where graph traversal techniques are general graph algorithms, \textit{shortest-path} algorithms discover the shortest path between nodes in a graph. One of the more famous algorithms in this category is the \textit{Dijkstra} algorithm. A popular variant uses a source node and calculates the shortest path to any other node in the graph. Common usages include navigating between two points on a map.

\subsubsection{Cs, Linear programming}

The \textit{linear programming} is a mathematical modeling technique where a linear function is either maximized or minimized under constraints. The behavioral outcome makes processes more efficient or economically cost-effective. This behavior means that any problem that requires more efficiency can take advantage of linear programming, whether the situation involves improving energy distribution or mathematical problem solving.


\subsection{Artificial intelligence, Ai}

\textit{Artificial intelligence} encompasses many algorithms, from object recognition (i.e., correlation) to the far-out attempts to create artificial life. We can crudely subdivide the subject into people-types \textit{connectionist}, \textit{evolutionist}, \textit{bayesian}, \textit{analogizer}, and \textit{symbolist}. In the early days, artificial intelligence covered everything we could not achieve. Today, a broader definition is used that defines the subject by problems. The general goal is to tackle evermore difficult challenges where the path is less well known.\par

As with Computer science, we do not cover all the algorithms in the field, but we will try to cover some interesting behavioral classes.


\subsubsection{Ai, Reinforcement learning}\label{sss:reinforcement}

\textit{Reinforcement learning} is an award-style algorithm. The algorithm rewards a path that gets closer to a solution. It encourages forward progression. The disadvantage is overlooking a better solution, a single-minded approach; nevertheless, it is a powerful algorithmic technique. The single-mindedness makes it potentially dangerous without some form of safeguards.

\subsubsection{Ai, Evolutionary algorithms}

Use a synthetic form of evolution to explore problem domains; if we consider the world as a two-dimensional graph with data on the x-axis and algorithms on the y-axis. Neural networks live near the x-axis, and evolutionary algorithms live near the y-axis. They modify algorithms. Either by playing with the variables or creating \& modifying the potential solutions directly. Similar to biological evolution, there is, for the most part, a population of potential solutions, and that population goes through generational changes. These changes occur through \textit{mutation} and \textit{crossover}. Each member of the population, in a generation, is valued by their \textit{fitness} towards the potential solution. A population can either start as an initial random seed (i.e., \textit{tabular rasa}) or with a known working solution. We use these algorithms for optimization and discovery. These algorithms are powerful, especially when the problem domain is too big or the solution is beyond human knowledge \cite{simon2013evolutionary}.

\subsubsection{Ai, Correlation}\label{ai:correlation}

\textit{Correlation} allows pattern recognition with a level of certainty. For example, \say{we are 87\% sure that the orange is behind the pineapple}. Neural nets provide certainty of recognition. \textit{Convolution neural networks} and \textit{deep learning} rely on this technique to solve problems. Hyperparameters configure the network, which is a complicated process. The network learns a response using training data. The quality of the training data determines the effectiveness of the correlation. This technique is good at image and speech recognition.

\subsubsection{Ai, Gradient descent}

\textit{Gradient descent} is about finding the best descent path down a steep hill. The technique reduces the cost and loss of setting up a neural network model by looking for the fastest descent. In other words, it is about finding the optimal minimum and avoiding the local minimum of a differentiable function. As the algorithm proceeds, it sets up the various parameters of a neural net model. Both Machine learning and, more specifically, deep learning use this technique.

\subsubsection{Ai, Ensemble learning}

\textit{Ensemble learning} uses multiple learning algorithms to obtain better predictive performance. Like predicting the weather, numerous futures are provided, from the extremes to the most likely. In other words, the ensemble method uses multiple learning algorithms to obtain better predictive performance than could be obtained from a single algorithm. An ensemble learning system consists of a set of different models: with diversity in structure (and hyperparameters). Its outward behavior is to produce several solutions in the hope of finding a better solution.

\subsubsection{Ai, Random forest}

\textit{Random forest} is a type of ensemble learning. Random forests apply to classification (the act of classifying due to shared properties), regression (finding cause and effect between variables), and tasks that involve building multiple \textit{decision trees} (decisions represented by a tree leading to a consequence). The random forests method generally outperforms traditional decision trees, but their accuracy can be lower than other methods.

\subsubsection{Ai, Continuous learning}

Continuous learning has a long history with traditional \textit{evolutionary algorithms}. An algorithm is left to learn in an environment with resources. The algorithm continuously adapts. Experiments have shown these algorithms exhibit the most significant learning jumps at the beginning of the cycle, and as time progresses, jumps become ever fewer, if not at all. This situation can alter if changes occur in the environment (e.g., additional resources or objectives).

\subsubsection{Ai, Novelty search}\label{ai:novelty}

\textit{Novelty search} is a different approach to, say, reinforcement Learning (see Section \ref{sss:reinforcement}), where the rewards are for learning new experiences rather than moving nearer to a goal \cite{stanley2015greatness}. Novelty search has the behavioral advantage of not requiring an initial objective for learning to occur. For example, learning to walk occurs by learning how to fall over. Falling over is not directly linked to the act of walking.

\subsubsection{Ai, Generative adversarial network, GAN}

\textit{Generative adversarial network} is, in fact, two networks \textemdash each one vying to attain different goals. One side is a creator (i.e., the generative network), and the other is a critic (i.e., the discriminative network). The creator's role is to outsmart the critic, meaning the creator can mimic the dataset. From a behavioral view, this algorithm can create a fake version (or deep fake output) of existing material. As of writing this text, this technique is taking over many traditional learning algorithms. 

\subsubsection{Ai, Supervised learning}\label{ai:supervised}

This class is a more generalized version of the correlation mentioned in Section \ref{ai:correlation}. \textit{Supervised learning} is a system using input-output pairs. In other words, known input samples connect to known outputs. The system learns how to connect the input to the output. The input data is labeled. Currently, the majority of machine learning is of this form. From a behavioral view, this algorithm requires \textit{human} supervision, as the name implies. The quality of training data is paramount. 

\subsubsection{Ai, Unsupervised learning}

\textit{Unsupervised learning} is the opposite of supervised learning (Section \ref{ai:supervised}). The input data is unlabeled, meaning it is not of a particular form. For example, no pictures labeled cats. This algorithm class attempts to find connections between the data and output. The unsupervised means no human has gone along labeling the data. From a behavioral view, this is a desired attribute but more challenging to control and get right \textemdash for example, the risk of connecting uninteresting features to an outcome. 

\subsubsection{Ai, Self-supervised learning}

It is a compromise between supervised and unsupervised learning. \textit{Self-supervised learning} learns from unlabeled sample data, similar to unsupervised learning. What makes it an in-between form is a two-step process. The first step initializes the network with pseudo-labels. The second step can be to use either supervised or unsupervised learning on the partially trained model from step one. From the behavioral view, not sure it shows any difference between supervised and unsupervised learning. 

\subsubsection{Ai, Bayesian probabilism}

\textit{Bayesian probabilism} is classical logic, as far as the known is concerned. New variables represent an unknown. Probabilism refers to the amount that remains \textit{unknown}. This class of algorithms is best for robotics, particularly Simultaneous Localization and Mapping (SLAM). These algorithms are good at mitigating multiple sources of information to establish the best-known consensus. For example, while a robot tracks its path, it always runs on a minimal amount of information and makes decisions based on statistical likelihood.

\subsubsection{Ai, Knowledge graphs}

As the name implies, \textit{knowledge graphs} use data models as graph structures. The graphs link information together. The information can be semantic (hold meaning). Knowledge-engine algorithms use knowledge graphs to provide question-answer-like services, i.e., expert systems. These algorithms, in theory, can provide some form of causality mapping since the graphs store the knowledge relationships. 

\subsubsection{Ai, Iterative deepening A* search}

Lastly for artificial intelligence, we decided to include one traditional artificial intelligence algorithm from the past. \textit{Iterative deepening A* search} is a graph traversal search algorithm to find the shortest path between a start node and a set of goal nodes in a weighted graph. It is a variant of the iterative deepening search since it is a depth-first search algorithm. We use these algorithms in simple game-playing \textemdash for example, tik-tac-toe or, potentially, chess.

\subsection{Quantum computing, Qc}

\begin{figure}[ht]
\begin{center}
\includegraphics[width=0.4\linewidth]{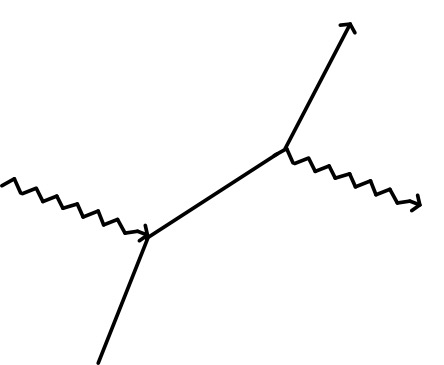}
\caption{Sum-over-histories (Feynman diagram)}
\label{figure:feynman}
\end{center}
\end{figure}

\textit{Quantum computing} is relatively new in the context of algorithms since hardware devices are rare and, if not difficult, at least different to program. When describing the world of quantum computing in a few paragraphs, it quickly becomes apparent that we could slide into an overly complex explanation. To avoid some of the complexity and remain relatively helpful, we decided to explain quantum computing from the perspective of how it differs to classical digital computing \cite{QC1}. Keep in mind quantum computers require a lot of classical computing to operate.\par

Quantum computers follow a different set of rules or principles. These rules come from atomic and subatomic particle physics, i.e., the notoriously complicated world of \textit{quantum mechanics} \cite{bhattacharya2022man,halpern2017quantum}. Classical digital computing uses transistors to implement bits. Quantum computers use even smaller atomic-scale elements to represent quantum bits, also known as qubits. A qubit is the unit of information for quantum computers. Transistors represent \underline{either} $0$ or $1$, binary qubits represent $0$ and $1$ simultaneously by including a continuous phase. Qubits, therefore, have the unique property of simultaneously being in a combination of all possible states. This fundamental principle of quantum mechanics is called \textit{superposition}. Superposition enables a quantum computer to have non-classical behavior. This non-classical behavior means we can probe many possibilities at the same time \cite{QC2} with the potential for saving considerable energy over classical computing. \par

Another principle used in quantum computers is \textit{entanglement}. The basic concept is that quantum systems can correlate so that a measured state of one can be used to predict the state of another. This connection enables the construction of quantum computers to include linkages between sets of qubits. It reinforces other physical principles, such as \textit{causality} limiting communications to no more than the \textit{speed of light}. Entanglement and superposition are intimately connected, in subtle ways, even across long distances. \par

Relative to a classical computer, a contemporary quantum computer has higher error rates, better data analysis capabilities, and continuous intermediate states. Compare this with classical computing, which has far lower error rates, is better for day-to-day processing, and uses discrete states \cite{QC1}. A quantum computer comprises a set of qubits, data transfer technology, and a set of classical computers. These classical computers initialize the system, control it, and read the results. Where the qubits carry out the computation, the transfer technology moves \textit{information} in and out of the machine. At the conclusion of the calculation, measurements of quantum states will return definite classical values at the outputs. Different runs on the same inputs return a distribution of results whose magnitudes squared are interpreted as a probability distribution. After completing the quantum calculation and measurements, classical computers do error correction, interpolation, and filtering across many runs of the same program. The information involves quantum (superposing and entangling qubits) and classical (configuring the initial values and reading out the classical final values). Modern quantum computers can efficiently collect statistics. These statistics provide more probable answers to more complex problems instead of definitive answers to simpler ones.\par

As with computer science and artificial intelligence, we will explore quantum algorithms, not from the quantum computing perspective but the algorithm perspective, i.e., \textit{what can they do?}. In keeping with the previous discussions, we provide a subset of algorithms. This technology has enormous potential but maybe ten or more years away. We believe it is essential to include this area when exploring future algorithms.

\subsubsection{Qc, Shor's algorithm}

\textit{Shor's algorithm} appears to be currently the most important, or most practical, in quantum computing. Shor's algorithm takes an integer $N$ and returns the prime factors. This is achieved in polynomial time (NP, see Section \ref{ss::computational}).  An alternative to \textit{Fourier transform}. Shor's algorithm uses the quantum Fourier transform to find factors. This behavior has implications for cryptography. It is an exponential speedup from the ability of a classical digital computer to break the types of cryptographic codes used for authentication and signatures (e.g., common methods such as RSA or ECC). This capability gives quantum computers the concerning potential to break today's security algorithms. \par

\textbf{Note}: In the press and academia, we now hear the term \textit{Post-Quantum Encryption} (PQE). PQE is a classical computing response to this capability. The response is a set of classical algorithms resistant to quantum methods. Many meta versions can exist because they can be dynamically updated and modified. Making the approach less reliant on the underlying hardware.

\subsubsection{Qc, Grover's algorithm}

\textit{Grover's algorithm} is probably the next most important quantum algorithm. Grover is said to be a quantum search algorithm. The algorithm carries out an unstructured search. And, as such is potentially useful to speed up database searches. There is a potential that classical algorithms could be equally competitive. The algorithm finds a unique input with the highest probability for a particular output value. This in theory can be achieved in $O({\sqrt{N})}$ time. Where $N$ is defined as the size of the function domain.

\subsubsection{Qc, Quantum annealing}

Quantum annealing (QA) is a \textit{metaheuristic}. A metaheuristic is a problem-independent algorithm normally operating at a higher level. Metaheuristics look at a set of strategies to determine the best approach. Quantum annealing finds a given objective function's extreme, either minimum or maximum, over a set of solutions. In other words, it finds the procedure that finds an absolute minimum for size, length, cost, or distance from a possibly sizable set of solutions. Quantum annealing is used mainly for problems where the search space is discrete with many extremes \textemdash limited only by available resources.

\subsubsection{Qc, Adiabatic quantum computation (AQC)}

Adiabatic quantum computation is reversible \cite{LANDAUER2002}. The word adiabatic means \textit{no heat transfer}, allowing for reversible computing, see Section \ref{sec:computation}. Calculations occur as a result of the adiabatic theorem. Optimization is the first application for these algorithms, but there are potentially many others. It is an alternative to the circuit model (from digital computing). This alternative makes it useful for both classical and quantum computation. 

\subsubsection{Qc, Quantum walks}

Lastly, \textit{Quantum walks} is the quantum equivalent of the classical random walk algorithm \cite{Venegas_Andraca_2012}. Similar to the other quantum solutions, a quantum walk operates with different rules. On certain graphs, quantum walks are faster and, by implication, more energy efficient than the classical equivalent \cite{KADIAN2021100419}.

\section{Hardware options}

\begin{quote} 
\centering 
\textit{"without hardware, we don't have algorithms, and without algorithms, there is no purpose for the hardware"}
\end{quote}

Even though we are mindful of analog solutions and the exciting developments in quantum hardware, we will focus primarily on digital solutions. We are also aware of Moore's Law's limitation, which may affect the future direction of computation, e.g., neuromorphic computing, DNA computing, analog computing, or unconventional computing. Maybe over time, there will be a change of emphasis toward analog, but today, digital systems lead. Digital systems include some form of traditional Turing machine. Turing machines are either fully implemented (e.g., a general-purpose processor moving towards the Principle of Universality) or partially implemented devices (e.g., a specialized accelerator missing some control elements).\par

Compute systems fall into two categories based on input data. The data is either embarrassingly helpful (i.e., sequential or parallel) or embarrassingly unhelpful. For the former, embarrassingly helpful, we design specialized hardware. For the latter, embarrassingly unhelpful, we design universal hardware to accommodate a broader range of problems. \par

\begin{table*}[ht]
    \begin{tabular}{lllll}
\hline
Year & Cause & Effect & & \\ \hline    
& & & & \\    
1912 & JK Flip flop  & Start of Boolean logic in circuits &  &  \\
1914 & Floating point  & Algorithms to handle real-world problems &  &  \\
1936 & Turing machine & Universal computation model &  &  \\
1943 & Finite automata &  Original pattern recognition method &  &  \\
1945 & ENIAC & First programmable computer, draft EDVAC report & & \\
1946 & Automatic Computing Engine (ACE) & RISC-style computation & & \\ 
1948 & Digital Signal Processing & Analysis and processing of continuous data  &  &  \\
1954 & SR, D, T & Adding temporal logic & & \\
1955 & Finite State Machine (FSM) & Complex pattern matching &  &  \\
1959 & Metal–Oxide–Semi. Field-Eff. Transistor (MOSFET) & Enabled far more sophisticated algorithms &  &  \\
1961 & Transistor–transistor logic (TTL) & Continuing to enable increased complexity &  &  \\
1961 & Virtual memory & Decoupling from physical constraints  &  &  \\
1964 & IBM System/360 & Inflection point in architectures & & \\
1965 & Memory Management Unit & Standard control of decoupling &  &  \\
1966 & Single Instruction, Multiple Data (SIMD) & Fast method of handling one dimensional arrays &  & \\
1967 & Virtualization & Make the underlying hardware virtual &  &  \\
1968 & IBM ACS-360 SMT & Full utilization of the processor & & \\
1971 & Intel 4004 & Allow for hard-coded algorithms (no stack) &  &  \\
1972 & Single Instruction, Multiple Threads (SIMT) & SIMD array processing algorithms & & \\
1972 & Packed SIMD & Speed-up software CODECs & & \\
1973 & Ethernet & Distributed (networked) algorithms & & \\
1975 & Dataflow & Execution flows on context  &  & \\ 
1976 & Harvard cache & Separating data and instruction efficiencies & & \\
1976 & RCA’s “Pixie” video chip GPU & Geometry based algorithms  & & \\
1978 & Ikonas RDS-3000 (claim first GPGPU) & Machine learning \& Crytocurrency & & \\
1979 & Motorola 68000 (CISC) & Execute sophisticated programming languages & & \\
1979 & Berkeley RISC & Change in algorithms to support load/store & & \\
1981 & Quantum computing & Probabilistic mathematics i.e., qubits & & \\
1983 & Networks (ARPANET) & Adoption of the TCP/IP protocol & & \\
1984 & SPMD & Messaging passing & & \\
1984 & VLIW & Instruction level parallelism & & \\
1985 & Intel 80386 & Mainstream MMU & & \\
1990 & Liquid Crystal Display & Ubiquitous flat screen monitors & & \\
1994 & Beowulf clusters & Using standard hardware for massive scale & & \\
1997 & Samsung DDR memory & Algorithms could be bigger and faster & & \\
2006 & MPMD & Games console (Sony PlayStation 3) & & \\
2013 & MIMD systems & Ubiquitous parallel compute nodes & & \\
2013 & Predicated SIMD & Associative processing & & \\
2020 & Unified memory & Heterogeneous compute sharing memory & & 
\end{tabular}
\caption{Cause-and-effect table for classical hardware (best guess)}
\label{table:history}
\end{table*}

It is challenging to map hardware developments to algorithm advancements, so we created the best guess using chronological ordering and the effects, see Table \ref{table:history}. We combined inflection points, and technology jumps to represent hardware advancements. This list is endlessly complicated, even when constrained. It is not clear when a technology caused an effect. If we look at history, what does it show us? It shows at least three emerging patterns: 

\begin{enumerate}[label=(\roman*),leftmargin=*]
    \item Long-term serial speed-ups have been the priority for hardware until relatively recently. In more modern times, we can see a steady increase in parallelism at all levels of the computation stack. Endless gains in parallelism from bit patterns to network scaling. Parallelism gives performance advantages for problems that can explicitly exploit such designs. The exploitation allows for more sophisticated algorithms and to scale out. 
    \item History of hardware has seen a constant struggle for and against the \textit{Principle of Universality}. In other words, a continuous fight between computing engines that can handle everything (\textit{Turing-complete}) and specialized ones (\textit{Turing-incomplete}).
    \item Sophistication of hardware has increased exponentially, allowing algorithms to be more capable and less efficient. Optimization is more complicated, and the likelihood of hitting an unusual anomaly is more likely.
\end{enumerate}

\section{Next algorithms}

\textit{What next for algorithms}? Making any prediction is difficult, but there are some standard frameworks we can apply. Firstly, we must determine where future algorithms will likely come from and why. To get this moving, we look at the existing algorithms; let us call this the $\alpha$ \textit{"alpha"} future. \par

The $\alpha$ future involves taking what we have and improving either the performance or efficiency of the algorithms. Conversely, we could also take an older algorithm and run it on modern hardware. Both these concepts are increments. These improvements should occur naturally and not necessarily cause a change apart from maybe more modularization and specialization. Algorithms that fit this category include \textit{fast Fourier transforms}, \textit{geometric mathematics}, and \textit{regular expression}. These algorithms rely on steady incremental improvements in both software and hardware. \par

Next, and remaining with the $\alpha$ future,  are the algorithms that initially start executing on much larger computer systems and eventually migrate over time to smaller systems. We predict that many of today's offline cloud algorithms, requiring specialized computation and vast resources, will ultimately become online algorithms on mobile devices (assuming some form of Moore's law still exists). For example, learning will move from the cloud to smaller devices in the coming decades. Learning algorithms are offline high-intensity applications, so processing does not occur in real time. A shift towards lighter real-time mobile variations will likely happen in the future, partly due to privacy concerns and partly due to economic ones (end users foot the bill). \par

The improvement in the  $\alpha$ algorithms allows a jump (not an increment) in new future algorithms. These new algorithms represent what we call the $\alpha\prime$ \textit{"alpha prime"} future. Many technological advances coincide, e.g., speech recognition, natural language processing, and gesture recognition. These technologies require improvements in existing algorithms and can be combined to help discover the new $\alpha\prime$ algorithms \cite{AImathematics,DEEPMIND}. For example, natural language processing will likely assist in creating future algorithms, i.e., moving away from programming languages toward negotiation, where we negotiate with existing ideas to produce a new solution. We need this to happen if we want to explore more of the natural world. It requires us to hide some development details, makes problem exploration more abstract, and leverages existing knowledge. \par

This abstraction means algorithm construction will likely change in that algorithms will be designed for re-purposing, modification, and interface negotiation. The modifiable part is to allow for multi-objective goals (see Section \ref{ss:objective}). Many next-generation algorithms expect insertion into bigger systems, where discovery and negotiation occur. Traditional methods are too rigid for the next set of problem domains. This change means algorithm development is more about setting the multi-objectives and goal-driven negotiation than gluing pieces of low-level code together. Below are some interface types that might be involved in an $\alpha\prime$ future:

\begin{itemize}[leftmargin=*]
    \item \textbf{No interface} - potentially entirely machine-driven, no human standard mechanisms are defined. In other words, an automation system works out its communication language. For example, Facebook proved this possible when two machine learning systems created their language for communication \cite{AILANG}.
    \item \textbf{Static interface} - human, and strict mathematical interface. Restricted to the lowest level if the compute node or pointers-to-memory provides basic types.
    \item \textbf{Dynamic interface} - again human and strict mathematical interface, similar in characteristics to static interfaces but not fixed to the image, i.e., late binding.  
    \item \textbf{Messaging passing} - human, less mathematical, unstructured, and non-standard, propriety concept. Powerful since it can handle heterogeneous systems. 
    \item \textbf{Messaging passing with ontological information} - human, semantic representation, structured so that the data can be discovered and assessed. Data is self-described over raw data.
    \item \textbf{Evolutionary interfacing}, where a method of evolution decides on a dynamic process of interfacing. The interface changes depending on workloads and a temporal element.
\end{itemize}

Lastly, we have the $\beta$ \textit{"beta"} future. The $\beta$ future for algorithms is more about the computing substrate. The substrate may change into an exotic computing zoo, e.g., quantum computing, DNA computing, optical computing, biological computing, and unconventional computing. Silicon will remain the digital workhorse, but the edge of algorithm development may shift. And with this change come very different approaches to algorithms and datatypes. It is exciting to look at new algorithms on these new computing options. \par

As well as the substrate, the types of algorithms in a $\beta$ future are different—for example, artificial general intelligence. Artificial general intelligence is currently a theoretical goal to create a universal intelligence that can solve many problems. A significant part of these algorithms is driving the decomposition (breaking a problem into smaller sub-problems) and recombination (taking the sub-problems and putting them back together to solve the main problem). Where the algorithms are much more general solutions than specialized ones, this is important as we try to handle problem domains at a much larger scale.

\subsection{Major meta-trends}

We see potentially three overarching meta-trends occurring in  $\alpha$, $\alpha\prime$ and potentially $\beta$ futures, namely \textit{parallelism}, \textit{probability}, and \textit{interaction}. Under those headings, we can link other subjects, such as artificial intelligence, quantum computing, and computer science.\par

\subsubsection{Parallelism}
\textit{Is there any more parallelism to be extracted}? We have taken what we call algorithmic \textit{structural parallelism} (e.g., data-level parallelism, instruction-level parallelism, and thread-level parallelism) to an extreme. Structural parallelism is when a problem breaks neatly into similar-looking sub-problems—covering embarrassingly sequential and parallel problem domains. But there are other forms of parallelism, for example, biological parallelism. Independent cells work together in parallel to form complex structures. Analogous to these natural processes are Carl Hewitt's concept of \textit{actors} \cite{hewitt2015actor} and John Hollands' view on complexity \cite{holland2014complexity}.\par

An actor is a small computational element comprising compute, memory, evolving rules, and adaptable communication (i.e., message based). Actors can have \textit{reason-response} capabilities. It moves away from mathematics and appears more like particle physics. Together with evolutionary techniques, actors can solve complex problems. Actors have to be free-flowing and can connect loosely; the loosely based connections are the reason for more message-passing style interfaces for communication. We have not fully utilized these technologies because (a) the success of classical parallel systems, (b) the lack of biological level scaling, and (c) advanced interfacing. Actors will likely play a much more important role in the future as more evolutionary technologies glue everything together. \par

\subsubsection{Probability}\label{ss:probability}

Computer science has long predicted the importance of probability. As we approach limitations in computation, uncertainty will start to dominate. We believe the subsequent algorithms will have to link directly or indirectly to probability. Whether quantum or classical computing, they all rely on statistical approximation over precision and accuracy.\par

For quantum computing, error correction will be paramount unless we can make them less noisy (highly unlikely). Error correction will most likely have to reside in classical computing. Even though quantum computing opens up new possibilities for algorithms in one direction, it causes problems in another (error correction). It is worth pointing out that voting systems (a class of Byzantine algorithms) are likely to become more common. Error correction codes are suitable for a specific problem, whereas voting systems are helpful for system-level corrections. \par

\subsubsection{Interaction}

\textit{Interaction} is a different take on future algorithms. \textit{What does the term interaction mean in this context?} It is to do with the process of creating new algorithms. This meta-trend brings new algorithms, old algorithms, and humans together \cite{AImathematics}. In other words, future systems can solve problems by selecting an old algorithm or creating a new one. This flexibility is possible by building advanced tools to explore the algorithm space, i.e., algorithms exploring algorithms. We are starting to see this with new programming languages, and libraries, allowing for greater expressiveness.\par

This greater expressiveness improves productivity by combining new tools with future user interface technologies (e.g., natural language processing, speech recognition, gesture recognition, patterning recognition, and goal-oriented design). We can see a meta-trend toward a more integrated algorithm exploration framework, moving away from implementation details. Transition is only made possible due to the advancements in digital hardware. \par

The interaction is to accommodate all the added complexity, vast data,  and navigation required to find a new algorithm. No longer can algorithm development occur without such advanced tooling. This becomes especially interesting if the tooling involved \textit{virtual reality} (VR), and \textit{augmented reality} (AR). \textit{Advancing mathematics by guiding human intuition with AI} \cite{AImathematics} by Davis et al., published in Nature in 2021, highlights that machine learning, with human assistance, is beginning to tackle hard mathematical problems. As algorithms become more expressive, they can be re-applied, with human assistance, to create even more algorithms. Solving previously intractable problems.\par

To provide a glimpse of future capabilities, let us look briefly at \textit{human intent} as part of a negotiated search for a new algorithm. Human intent is essential for many technology areas since it is about second-guessing what a human intends to do. This guessing may become one of algorithm development's most potent tools. Allowing a computer system to understand the intent of a human objective. \par

\subsection{Other trends in algorithms}

In this section, we discuss some of the other potential trends in algorithms, some are just a continuation of existing trends, but others are emerging trends that may become important.

\begin{enumerate}[label={OT-\arabic*.},leftmargin=*]
    \item \textbf{\textit{Automation of everything}}, algorithms continue to automate activities at every level \textemdash, for example, the automation level required to process package distribution within a warehouse.
    
    \item \textbf{\textit{Growth of exploration over exploitation}}, exploitation remains in common practice, but the use of exploration is rising. We want to understand more about the natural world. This change will increasingly occur as we shift to problems beyond human capability.
    
    \item \textbf{\textit{Mathematics becomes a target}}. Algorithms can optimize older equations and formulas \cite{DEEPMIND}. Algorithms can search a much bigger problem domain than any human in the past \cite{DEEPMIND}. We will see the optimization of traditional mathematics using modern techniques. A future mathematician will most likely be very different from the past.
    
    \item \textbf{\textit{Spiking neurons}},  basically any algorithms that allow backward information feed, i.e., from a higher level back to a lower level, improve recognition or optimization. This biological method can potentially reduce the size of the required networks and, in many cases, could enhance the quality of results (connection to neuromorphic computing).
    
    \item \textbf{\textit{Development through negotiation}}, already mentioned but worth repeating is the creation of algorithms that allow humans to define goals. The goal is a starting point for navigating complex problem domains. Solutions are developed over time through negotiation \cite{BretVictor}.
    
    \item \textbf{\textit{Pressure on better knowledge representation}}. Knowledge representation is at the core of all the activities around algorithms. The pressure is due to the problem domains becoming more complicated. This trend will continue, and we will likely see an expansion of basic data types.
    
    \item \textbf{\textit{Pattern recognition through increased dimensionality}}. With increasing resources, adding data dimensions will continue as a style of pattern extraction from complex and noisy data.
    
    \item \textbf{\textit{Single shot learning algorithms}}, is the ability to learn with minimum data sets. We see a continuous trend to reduce the data required to train the next-generation algorithms. 
    
    \item \textbf{\textit{Sparse data structures}} will continue to be important. Using these structures is a desperate attempt to reduce resource requirements and improve performance. This measure becomes especially important for algorithms that require enormous data sets. 
    
    \item \textbf{\textit{Prediction}}, we can say that predictive model will occur in every critical area, from social decision-making to instruction pre-fetchers. This trend continues unabated for the foreseeable future. 
    
    \item \textbf{\textit{Physical three-dimensional algorithms}}, these are algorithms that deal with the layout and positioning of physical components. Important for 3D transistor layout, virtual reality systems escaping the real world, and augmented reality systems that add to the real world. 
    
    \item \textbf{\textit{Map between physical and virtual worlds}} will increase in importance. This mapping is required if we want to accelerate the adoption of simulation, i.e., transferring environments quickly into virtual representations.
    
    \item \textbf{\textit{Byzantine algorithms}} become more critical as society deploys machine learning models. Multiple model voting increases the likelihood of a correct prediction; machine learning will move quickly in this direction to avoid biases. 
 
    \item \textbf{\textit{Generative Adversarial Network (GAN)}} continues to be more successful and valuable along with traditional machine learning systems.
    
    \item \textbf{\textit{Built-in multi-objective capability}}. These options allow for more of a weather prediction-type approach to solutions. The variations can range from maximum optimization to zero optimization or likely to most unlikely; see Figure \ref{figure:pareto}.
    
\end{enumerate}

\section{Auxiliary support}

We have created a list of what might be next for algorithms, but what about auxiliary support?

\begin{figure*}[ht!]
\begin{center}
\includegraphics[width=0.7\linewidth]{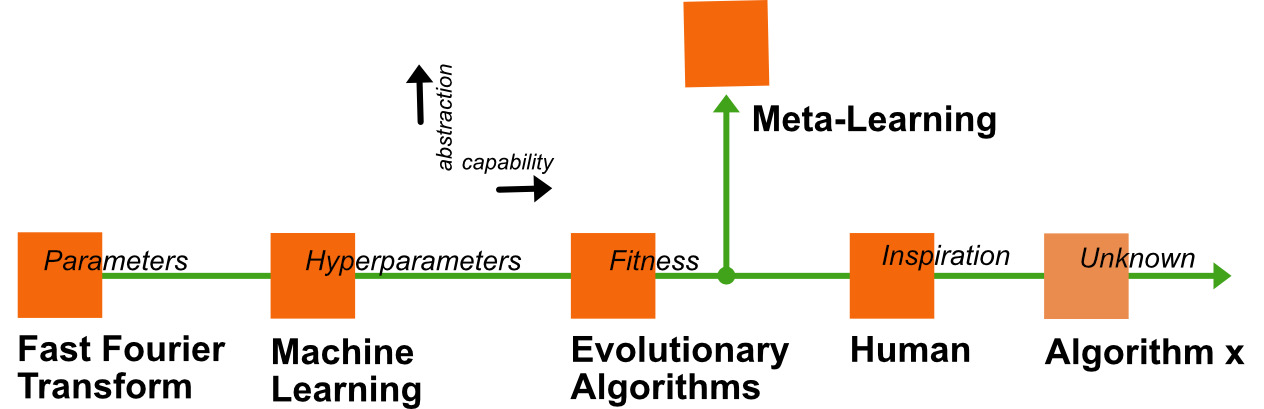}
\caption{Meta-Learning}
\label{figure:inter}
\end{center}
\end{figure*}

\begin{enumerate}[label={HG-\arabic*.}]

    \item \textbf{\textit{Multi-level randomness}}. The requirement for much more sophisticated randomness. Having multiple levels of \textit{good} randomness becomes imperative. What we mean by good randomness is having everything from, as near as possible, true randomness to pseudo-randomness. We require an ability to dial randomness to a particular level, including some form of stochastic numbers. 
    
    \item \textbf{\textit{Probabilistic operators}}, extension of operations to include probabilistic helper functions. We have integer, fixed point, and floating point operations. We need probabilistic operators. 

    \item \textbf{P-adic datatypes and operators}, we add this as a potential auxiliary extension. p-adic extends the standard number systems adopted by digital computers \cite{TIMP2022,https://doi.org/10.48550/arxiv.1911.09624}. Based on prime numbers, p-adic allows for a different style of flexibility from the traditional extension of real and complex numbers. 
  
    \item \textbf{Stochastic rounding}, is already gaining momentum. It rounds real numbers using a probability distribution compared to the traditional fixed rounding up or rounding down to the nearest number \cite{TIMP2022,doi:10.1137/20M1334796}. This method is increasing in popularity, especially with the machine learning community, opening the door to lower precision numbers. 
    
    \item \textbf{\textit{Biological neuron mimicking}}. If we compare artificial and biological neurons, the biological neurons have many more links. We predict a change in the base neuron for future machine learning.
    
    \item \textbf{\textit{Memory management optimization}}, as memory systems become more sophisticated and complicated, we need new methods to help algorithms optimize memory efficiency and usage. This help may reside in hardware or future tooling. 
    
    \item \textbf{\textit{Agent or chaotic based parallelism}}, as mentioned previously, structural parallelism continues at all levels. Still, there is a potential for a hardware-assisted agent or chaotic, based parallelisms.
    
    \item \textbf{\textit{Error-correction}}, is an old subject with a new set of focuses coming from critical areas such as quantum computing, probabilistic systems, and traditional digital systems where geometry shrinkage goes to the limits. Any system that operates at the boundaries of stability where minor errors can result in significant problems \cite{HAMMING}.  

    \item \textbf{\textit{Spatial-temporal datatypes}}, as we move into more graphing problems (e.g., Virtual Reality, Augmented Reality, Digital Twin, and Physical Simulators), there is a requirement to make spatial-temporal datatypes a first-class citizen. A universal datatype for physical systems with potential sparse recursive scaling coordinate systems and velocity coordinates to represent \textit{n-body} problems \cite{TIMP2022}.
    
\end{enumerate}

\section{Conclusion}

We have taken a journey through the algorithm world. It has involved crawling through tunnels, jumping over fences, and running across fields. As with many subjects, we picked a few tunnels, fences, and fields, realizing this is a staggeringly small subset of the ideas. The algorithm world is complex, dynamic, and full of old and exciting new directions. As we see it today, there is an undertone that probability and statistics have an increasingly critical role in tackling complex applications. These problem solutions are less amenable to simple absolutes. \par

Over time, the application focus has shifted from calculating the ordnance range for artillery to pattern correlation using machine learning. These shifts have transitioned the algorithm world in different directions. We see, through our journey, new transitions towards data-directed outcomes, adaptability, and meta-learning.\par

 \textbf{\textit{Data-directed outcomes} replace \textit{rule-based systems}}. This transition is not new; we see this as a continuation. Rule-based systems can handle specific problem domains, but they fail when a none pre-programmed pattern occurs, i.e., they lack flexibility. Data-directed outcomes can circumvent, within reason, many of these problems, which are difficult for rule-based systems. For rule-based systems, the value often is in the exceptions, not the rules, and for data-directed outcomes, the value comes down to the quality of the training data. We may see a mixture of the two systems, with rule-based systems ensuring the other operates within required boundaries. \par

\textbf{\textit{Adaptability} replaces \textit{precision}}. Precision deals with perfection, whereas adaptability is handling imperfection. For example, we design robots to perfect specifications. And algorithms rely on those perfections for length, pressure, and movement. Adaptability in robots has algorithms that constantly learn and change as the robot matures or the environment changes. In other words, the new algorithms handle environmental changes, wear, and poor design. \par

\textbf{\textit{Meta-learning} enhances \textit{capability}}. Increasing capability is important; it is a horizontal activity. See Figure \ref{figure:inter}. We want to expand algorithm capability, so they tackle evermore exciting tasks. At the same time, we also want to accelerate the actual learning and creative process. Meta-learning is the process of \textit{learning to learn}, moving away from the specific problems and focusing on common patterns. These generalizations force algorithms up the abstract tree. In that, common patterns can transfer to other problem types. In the coming decades, we predict much more activity around meta-learning and integrating more abstract approaches. In addition, this could mean more weather-predicting style algorithms, i.e., ensemble prediction, that provide a range of solutions with different characteristics. In other words, a group of solutions with different probabilities of certainty and uncertainty \cite{TIMP2022,JamesAng}. We can potentially build these systems using multiple models based on physics and probability. It allows us to explore the unlikely so we can create anticipating actions with cost-loss attributes. For example, moving people out of danger to avoid a unlikely but possible catastrophic weather system \cite{TIMP2022}.  \par

In any modern algorithm discussion, it is essential to mention quantum algorithms. The quantum world is attractive for its potential energy saving advantages \cite{LANDAUER2002}. We are still in an exploratory phase and starting to learn how to build basic systems and determine possible algorithms. One of the many concerns about quantum computing is the quick drive to optimization before the benefits are genuinely discovered, i.e., the race to be valuable. Also, the problem domains that quantum computing can explore may not be that exciting, and traditional computation may remain dominant for the majority. One true unarguable benefit of quantum computing is exploring the quantum world itself \cite{UNCONV}. Quantum algorithms' ultimate achievement may be to push classical computing in new directions. \par

\textit{Hic sunt dracones} (\textit{\say{Here be dragons}}) is the term used at the beginning of this exploration. The term describes the world beyond the edge of the map. We are entering an exciting time around algorithms and what they can accomplish, but concerns about how they can be abused or used for badness come with that excitement. For example, we have seen algorithms manipulate people on a mass scale through social media. Or the various fake images and videos depicting people saying fictional opposites or non-truths. Just the mechanical process of \textit{validation} and \textit{verification} becomes more of a challenge as algorithms exhibit more extraordinary capabilities.\par\par

We want algorithms to be benevolent in our society. We have seen how algorithms can influence people away from acting in their best interests. For this reason, we provided a list of ideals at the beginning of our journey. These ideals are possible areas of further exploration, but they are not rules. At best, guidelines.\par

Lastly, \textit{Richard Hamming}, \textit{Albert Einstein}, \textit{Neil deGrasse Tyson}, and many others pointed out a common mistake: assuming the new is just like the past. The mistake prevents many from contributing significantly to the next revolution. For Hamming, this thought came as he observed the transition to digital filters \cite{HAMMING}.
 
\section{Acknowledgments}

The OPEN DASKALOS PROJECT is an open project to help explain complicated Computer Science concepts. Each paper is open and undergoes a continuous process of refinement, i.e., a snapshot in thinking. We thank the great algorithm writers for creating such exceptional solutions. We would also like to thank the reviewers, Hannah Peeler, for the initial feedback and Paul Gleichauf for posing such hard questions throughout the editing process.

\section*{OPEN DASKALOS PROJECT series:}

\textbf{Intelligence Primer (2nd Edition)}, May 2022\\
by Karl Fezer and Andrew Sloss\\\\
\textit{Intelligence is a fundamental part of all living things, as well as the foundation for \textbf{Artificial Intelligence}. In this primer we explore the ideas associated with intelligence and, by doing so, understand the implications and constraints and potentially outline the capabilities of future systems. Artificial Intelligence, in the form of Machine Learning, has already had a significant impact on our lives.} \\

\noindent
URL: \url{https://arxiv.org/abs/2008.07324}

\bibliographystyle{ACM-Reference-Format}
\bibliography{lost-in-algorithm}

\end{document}